\documentclass[twocolumn,prd,superscriptaddress,showpacs,showkeys,10pt]{revtex4}

\usepackage{amsmath}            

\usepackage{graphicx}           
\usepackage{dcolumn}            
\usepackage{amssymb}            
\usepackage{wasysym}            
\usepackage{hypernat}           
\usepackage{hyperref}           

\newlength{\defaultfig}         
\newlength{\widefig}            
\newlength{\narrowfig}          

\newcommand{\half}{{\textstyle \frac{1}{2}}} 

\graphicspath{{figures/}}       

\begin{document}

\title{Interactions of multi-quark states in the
  chromodielectric model}

\date{\today}

\author{Gunnar Martens}
\email[e-mail: ]{martens@th.physik.uni-frankfurt.de}
\author{Carsten Greiner}
\affiliation{Institut f\"ur Theoretische Physik, 
  Universit\"at Frankfurt, Germany}
\author{Stefan Leupold}
\author{Ulrich Mosel}
\affiliation{Institut f\"ur Theoretische Physik, 
  Universit\"at Giessen, Germany}

\begin{abstract}
  We investigate 4-quark ($qq\bar{q}\bar{q}$) systems as well as
  multi-quark states with a large number of quarks and anti-quarks
  using the   chromodielectric model. In the former type of systems
  the flux distribution and the 
  corresponding energy of such systems for planar and non-planar
  geometries are studied. From the comparison to the case of two
  independent $q\bar{q}$-strings we deduce the interaction potential
  between two strings. We find an attraction between strings
  and a characteristic string flip if there are two degenerate
  string combinations between the four particles. The interaction
  shows no strong Van-der-Waals forces and the long range behavior of
  the potential is well described by a Yukawa potential, which might
  be confirmed in future lattice calculations. The multi-quark states
  develop an inhomogeneous porous structure even for 
  particle densities large compared to nuclear matter constituent quark
  densities. We present first results of the dependence of  
  the system on the particle density pointing towards a percolation
  type of transition from a hadronic matter phase to a quark matter
  phase. The critical energy density  is found at $\varepsilon_c = 1.2 \,
  \text{GeV}/\text{fm}^3$. 
\end{abstract}



\maketitle


\section{Introduction}
\label{sec:intro}


\setlength{\defaultfig}{0.9\columnwidth}
\setlength{\widefig}{\textwidth}
\setlength{\narrowfig}{0.9\columnwidth}

Quantum Chromodynamics (QCD) is the theory for color charged quarks
and gluons. This theory has been tested successfully in the regime of
large momentum transfer such as in deep inelastic scattering, where
perturbative methods can be used due to asymptotic freedom. It is a
common belief that QCD should be able to describe all systems
ruled by strong interactions. These cover a wealth of different
regimes ranging from the dynamics of quasi free quarks and
gluons in a quark gluon plasma (QGP) at high temperatures or
densities, over the formation of hadrons out of quarks to the
interactions between those color neutral hadrons. However, in this
latter region of small relative momenta,
where the confinement phenomenon plays a dominant role, no convincing
analytical techniques have been established yet. The only calculations from
first principles are restricted to QCD lattice simulations, where the
results are limited due to today's computing power.

Therefore one still has to rely on models, that describe the
interactions between quarks and gluons bound to hadrons
phenomenologically and that include confinement. Such models are
for example the well known MIT bag model
\cite{Chodos:1974je,Chodos:1974pn}, where quarks are confined by a
given external cavity or the quark molecular dynamics model,
\cite{Hofmann:1999jx,Scherer:2001ap,Akimura:2004ec}, where
quarks follow the Hamiltonian dynamics subject to a  linear rising
potential between quarks ($q$) and anti-quarks ($\bar{q}$). Another
approach is the model of the dual super conductor known also dual
Abelian Higgs model or dual Ginzburg-Landau model
\cite{Baker:1991bc,Ripka04:dual_super} where confinement is achieved
by monopole condensation \cite{'tHooft:1974qc,Polyakov:1974ek} and an
accompanying magnetic supercurrent. The model of the stochastic vacuum
relies on the calculation of Wilson loops in a Gaussian approximation
\cite{Dosch:1987sk,Kuzmenko:2000rt,Shoshi:2002rd} which leads to a
linearly rising $q\bar{q}$-potential. 

In this work we choose the framework of  the chromodielectric model
(CDM) \cite{Friedberg:1977eg,Friedberg:1977xf,wilets:1989}. Unlike the
MIT bag model, CDM has the 
benefit, that bags with a smooth surface are created self-consistently
and dynamically out of the  underlying field equations, due to the
presence of colored quarks. 

The CDM has already been used to calculate properties of the
nucleon and its low lying resonances like masses, magnetic moments and
the axial-vector/vector coupling constant ratio \cite{Goldflam:1982tg}. 
In \cite{Bickeboeller:1985xa,Grabiak:1990yf,Martens:2003yy} a
description of $q\bar{q}$-strings was given, and in
\cite{Martens:2004ad} the parameters of the CDM were adjusted to
reproduce results of lattice calculations
\cite{Bali:1995de,Bali:1997am}. In the same work, the flux tube
structure of a baryon like $qqq$-bag has been studied.
Further, within the model, the interactions between $q\bar{q}$-strings
\cite{Loh:1997sk} and the nucleon-nucleon interaction in vacuum
and in nuclear matter
\cite{Pirner:1984hd,Achtzehnter:1985ur,Schuh:1986mi,Koepf:1994uq,Pepin:1996mp} 
has been discussed.  
In another approach the model has been used in a transport theoretical
framework to describe the dynamics of quarks bound in nucleons and
strings \cite{Kalmbach:1993sp,Vetter:1995gp}. In \cite{Loh:1997cr} the
disintegration of $q\bar{q}$-pairs in strong color electric fields
has been observed. A full molecular dynamical simulation of
hadronization out of a gas of quarks and gluons has been presented in
\cite{Traxler:1998bk}. 

In this paper we will analyze the interactions between the color
electric flux tubes for a wide class of different quark
configurations. In a previous paper \cite{Martens:2004ad} we have
given the structure of meson like $q\bar{q}$-states and baryon like
$qqq$ states. The model has been successfully adjusted to reproduce
lattice results for both the $q\bar{q}$-potential and the transverse
shape of the flux tubes. It is an open question, how the flux tubes of
the basic white two- and three-particle clusters interact with each
other. In order to understand the interactions we extend our previous
analysis to configurations with more than three particles. The easiest
of such systems, the $qq\bar{q}\bar{q}$-system, already develops two
distinct bags, that may interact with each other. We can study those
systems for different spatial quark configurations. Besides this,
the system is still simple enough to be treated on the lattice. In
fact there are lattice results for the four-particle system in
SU(2) \cite{Green:1993ag} and also in SU(3) \cite{Okiharu:2005ab}. 
Where possible, we will compare our results with those
obtained on the lattice. Knowing the interactions between the flux
tubes of color neutral objects, it is an interesting issue, how this
interaction governs the transition from a system of distinct white
hadrons to a system of interacting colored quarks in a quark
plasma. It is an old prediction of QCD, that there is a rapid
transition for increasing temperature \cite{Karsch:2000ps}. This
transition to the quark gluon plasma (QGP) has been 
explicitly seen for vanishing baryon chemical potential as a steep
rise of the thermodynamic pressure of the system at a temperature
$T=(170\text{-}180)\,$MeV \cite{Karsch:2000kv}. For non-zero
baryo-chemical potential, lattice 
calculations suffer from technical problems and cannot be easily
performed up to now. See \cite{Fodor:2004mf,Fodor:2004nz},
where the transition temperature was explored for small chemical
potential. However, for increasing baryon densities one expects that
the hadrons start to overlap and the quarks are free over a much larger
volume and a transition to a quark gas might occur as well. In
contrast, in our treatment of the CDM we have 
only static configurations and therefore we do not describe quark
systems at non-zero temperature. But it is possible to vary the quark
density and study many-quark systems and the behavior of the very
dense flux tubes.

The paper is organized in the following way. In section~\ref{sec:cdm}
we introduce briefly the chromodielectric model, give the equations of
motion for the underlying fields as well as the corresponding field
energies. We solve these equations numerically in three spatial
dimensions. We also discuss the color structure of the model and its
connection to the SU(3) color algebra. 
Section \ref{sec:long-range} is devoted to the long range
behavior of fields of $q\bar{q}$-strings, which extends the discussion
of the bulk properties of strings in \cite{Martens:2004ad}. We will
show the characteristic exponential decay of all fields with the
distance from the string center.
In section \ref{sec:stringint}
we show the interactions between two $q\bar{q}$-strings of various
lengths and with relative orientations to each other.
We first show our results for the color electric fields in section
\ref{sec:fields} and the distribution of the energy density in section
\ref{sec:distribution}. The 
distortion of the one $q\bar{q}$-string under the influence of the other
one is shown explicitly. For simplicity, we concentrate 
on the discussion of strings that are parallel or anti-parallel to
each other. In this case, all particles are lying in a single plane.
However, as the calculations are done in three dimensions, we are able
to study particle configurations too, that are extended in three space
dimensions. We will show the results of the calculations for these
tetrahedronic configurations as well. In section
\ref{sec:potentials} we give the interaction potential of the two
strings as calculated from field energies. Again we concentrate on the
potential for plane configurations but we will discuss also a wide
range of 3-dimensional configurations. We will extract the static
string interaction potential which can be compared to lattice
calculations. 
In section \ref{sec:multi-quark} we present first results on
static multi-particle systems. We show the heterogeneous structure of
the emerging flux tubes and analyze the scaling of the multi-quark
properties with the particle density. A qualitative description of the
deconfinement phase transition is given. Finally we discuss our
results in  section \ref{sec:summary} and give our summary. 

\section{The chromodielectric model}
\label{sec:cdm}

As there is no confinement in quantum electrodynamics, one believes
that confinement is due to the non-Abelian nature of QCD. Although
the detailed mechanism of confinement has not been revealed within
QCD, there is strong evidence for it from lattice QCD. The 
most prominent result is the linear rising potential between a static
quark anti-quark pair at zero temperature
\cite{Bali:1992ab,Deldar:1999vi} and at finite temperature
\cite{Bornyakov:2004yg}. In addition, the formation 
of long, string like flux tubes between a $q\bar{q}$-pair has been
seen both in SU(3) \cite{Matsubara:1994nq,Ichie:2002dy} and in SU(2)
\cite{Bali:1995de}, where the results are much more precise for
numerical reasons. Despite the fact that the structure of the QCD
vacuum is highly complicated, its long range behavior might be
transparent. In the CDM it is assumed that this
vacuum behaves as a perfect dielectric medium described by a vanishing
dielectric constant $\kappa_\text{vac}$. Dual to a normal
superconductor, where magnetic fields are expelled from the
superconducting phase, in CDM the color electric fields are expelled
from the QCD vacuum. In the presence of color charged quarks, the
resulting color fields are compressed into flux tube like excitations
of the vacuum. In \cite{Mack:1983yi,Pirner:1984hd,Pirner:1992im} a 
renormalization group derivation was given for the lattice
colordielectric model, which has a scalar confinement field, and
keeps strongly coupled non-Abelian fields in the large distance limit.
CDM is formulated in terms of two Abelian color fields $A^{\mu,a}$
only and an additional scalar confinement field $\sigma$. This scalar
field is designed to already include the non-Abelian effects of the
gluon sector. There are  
indications that in the  long range limit only the Abelian components
contribute to the observable quantities: 't Hooft suggested in
\cite{'tHooft:1974qc} the maximal Abelian gauge for projecting out a
Cartan subgroup believed to be relevant for infrared aspects of
QCD. In \cite{Ezawa:1982bf,Kronfeld:1987vd,Amemiya:1998jz} further
support for the Abelian dominance was found due to the mass generation
of off-diagonal gluons. In \cite{Shiba:1994ab,Bali:1996dm} the
string tension in the Abelian approximation was reproduced within some
percent deviation from the full value. 

Following this reasoning, CDM
is formulated in a Cartan subgroup of QCD reducing the independent
color fields to a set of two commuting field $A^{\mu,a}$ with $a \in
\{3,8\}$. These two fields are connected to the Gell-Mann matrices
$\lambda^{3,8}$ which commute with each other in the standard
representation. Confining effects are merged into the dielectric
coupling of these gluon fields to the dielectric medium generated by
the confinement field $\sigma$. The CDM Lagrange density can now be
given as
\begin{subequations}
  \label{eq:lagrangian}
  \begin{eqnarray}
    \label{subeq:lagrangean}
    {\cal L} &=& {\cal L}_g + {\cal L}_{\sigma} \quad ,\\
    \label{subeq:gluon}
    {\cal L}_g &=& - {\textstyle \frac{1}{4}} {\kappa(\sigma)}
    F_{\mu\nu}^{a} F^{\mu\nu, {a}}-g_s \;j_\mu^a \,A^{\mu,a} \quad , \\
    \label{subeq:sigma}
    {\cal L}_{\sigma} &=& {{\textstyle \frac{1}{2}}
      \partial_\mu\sigma\partial^\mu\sigma  - U(\sigma) } \quad , \\
    \label{subeq:gluontensor}
    F^{\mu\nu, { a}} &=& \partial^\mu A^{\nu,a} 
    - \partial^\nu A^{\mu,a}, \;\;\; a \in\{3,8\}  \quad ,
  \end{eqnarray}  
\end{subequations}
where $A^{\mu,a} = (\phi^a, \vec{A}^a)$ is composed out of the color
electrostatic potential $\phi^a$ and the vector potential $\vec{A}^a$.

In the following we are interested especially in the interactions of the
electric flux tubes, thus we have omitted the dynamic term for the quark
degrees of freedom. Quarks enter into  the  model only via the
external color current $j^{\mu,a} = (\rho^a,\vec{\jmath}^{\,a})$ with a
coupling strength $g_s$. Furthermore we treat
the quarks as infinitely heavy, static sources. Therefore the color
current vanishes $\vec{\jmath}^{\,a} = 0$. The color charge density
$\rho^a(\vec{x}) = \sum_k q_k^a w(\vec{x}-\vec{x}_k)$ is given as a
sum over all quarks with two color charges $q^a$ and a spatial distribution
$w(\vec{x})$. In principle the quarks are point-like objects, but for
numerical reasons we assign a Gaussian distribution  $w(\vec{x}) = (2\pi
r_0^2)^{-3/2} \exp(-\vec{x}^2/2 r_0^2)$ with a small width
$r_0=0.02\,\text{fm}$, which is on the order of the spacing of the
numerical grid used in the calculations. The grid spacing chosen is
small compared to all physical sizes like the flux tube radius.

\begin{figure}[htbp]
  \begin{center}    
    \includegraphics[width=\defaultfig,keepaspectratio,clip]{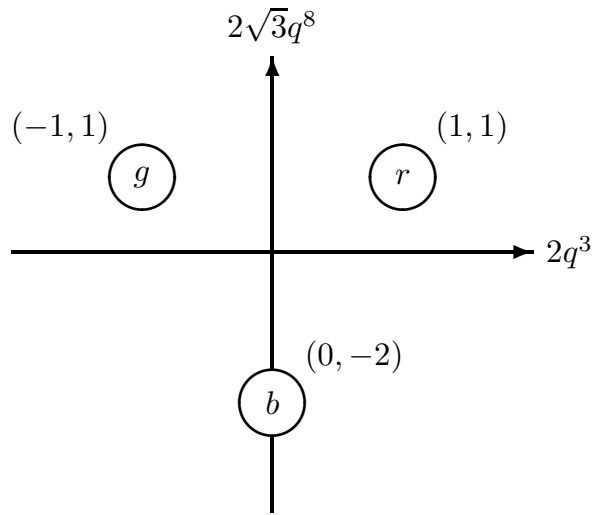}
    \caption{The color charge $(q^3,2\sqrt{3}q^8)$ with respect to the color
      fields $A^{3,8}$ for the three colors $r,g,b$.} 
    \label{fig:triplett}
  \end{center}
\end{figure}

In the Abelian approximation the quarks still have three different
colors. These are expressed as three dimensional unit vectors in the
fundamental representation of color space as $|c\rangle \in
\{|r\rangle,|g\rangle,|b\rangle\}$. The color charges are then given by
the diagonal entries of the corresponding generators
$t^a=\lambda^a/2$, i.e.~$q^a=\langle c| t^a |c \rangle$. Formulated
differently, the two color charges $(q^3,q^8)$ are given by the weight
vectors of QCD \cite{Scherer:2001ap}. The numerical values of the color
charges can be read off from table \ref{tab:charges} and are depicted
in figure \ref{fig:triplett}.

\begin{table}[htbp]
  \begin{ruledtabular}
    \begin{center}
      \begin{tabular}{lrr}
        color    &   $q^3$   &   $q^8$ \\\hline
        red      &   $1/2$     &   $1/(2\sqrt{3})$ \\
        green    &  $-1/2$     &   $1/(2\sqrt{3})$ \\
        blue     &   $0$     &   $-1/\sqrt{3}$
      \end{tabular}
      \caption{The color charges $q^a$ of the three colors with respect
        to the two Abelian color fields. The charge of the anti-colors
        are given by the negative of the color charges.}
      \label{tab:charges}
    \end{center}
  \end{ruledtabular}
\end{table}

The model inherits a U(1)$\times$U(1) gauge symmetry and a global
symmetry corresponding to finite rotations in color space
\cite{Martens:2004ad} which is a remnant of the original SU(3)
symmetry of QCD. Only the Lagrange density and the energy density, 
which is stated explicitly later on, are invariant under these
rotations but not the color fields themselves.

The color field tensor $F^{\mu\nu,a}$ ($a\in\{3,8\}$) in
Eq.~\eqref{subeq:gluontensor} defines the color electric and 
magnetic fields $\vec{E}_i^a = - F^{0i,a} = \left(-\nabla
  \phi^a-\partial_t \vec{A}^a\right)_i$ and $\vec{B}_i^a = -\half
\varepsilon_{ijk} F^{jk,a} = (\nabla \times \vec{A}^a)_i$
respectively. With the definition of the medium fields $\vec{D}^a =
\kappa(\sigma) \vec{E}^a$ and $\vec{H}^a =\kappa(\sigma) \vec{B}^a$
the equations of motion following from Eq.~\eqref{eq:lagrangian} are
given by
\begin{subequations}
  \label{eq:maxwell}
  \begin{eqnarray}
    \label{subeq:gauss}
    \nabla \cdot \vec{D}^a &=& g_s \rho^a \quad ,  \\
    \label{subeq:ampere}
    \nabla \times \vec{H}^a - \partial_t \vec{D}^a &=& g_s
    \vec{\jmath}^{\;a} \quad , \\ 
    \label{subeq:faraday}
    \nabla \times \vec{E}^a + \partial_t \vec{B}^a &=& 0 \quad ,\\
    \label{subeq:nomonopoles}
    \nabla \cdot \vec{B} &=& 0 \quad ,\\
    \label{subeq:eom_scalar}
    \left(\partial_t^2 - \vec{\nabla}^2\right) \sigma 
    + U'(\sigma) &=&  - {\textstyle \frac{1}{4}} 
    \kappa'(\sigma) F_{\mu\nu}^a F^{\mu\nu,a} \quad .
  \end{eqnarray}
\end{subequations}
With $\vec{\jmath}^{\;a}=0$ and the assumptions that all time
derivatives vanish exactly and $\vec{A}^a=0$, the magnetic field is
$\vec{B}^a = 0$ and the 
equations of motion (e.o.m.) can be cast into the following form:
\begin{subequations}
  \label{eq:static_eoms}
  \begin{eqnarray}
    \label{eq:poisson}
    \nabla\cdot(\kappa(\sigma) \nabla\phi^a) &=& -g_s \rho^a \quad ,\\
    \label{eq:static_sigma}
    \nabla^2 \sigma - U'(\sigma) &=& - \frac{1}{2} 
    \frac{\kappa'(\sigma)}{\kappa^2(\sigma)}
    \vec{D}^a\cdot\vec{D}^a \quad .
  \end{eqnarray}
\end{subequations}
The energy of the system for static configurations is given by 
\begin{subequations}
  \label{eq:energy_decomp}
  \begin{eqnarray}
    \label{subeq:energy_1}
    E_\text{tot} &=& E_\text{el} + E_\text{vol} + E_\text{sur} \quad ,\\
    \label{subeq:energy_el}
    E_\text{el} &=& \half \int d^3r \, \vec{E}^a\cdot\vec{D}^a \quad ,\\
    \label{subeq:energy_vol}
    E_\text{vol} &=& \int d^3r\, U(\sigma) \quad ,\\
    \label{subeq:energy_sur}
    E_\text{sur} &=& \half \int d^3r \, \left(\nabla \sigma\right)^2
    \quad ,
  \end{eqnarray}  
\end{subequations}
where we have labeled the different energy parts as total energy,
electric energy, volume energy, and surface energy, respectively, as
explained in \cite{Martens:2004ad}. 

\begin{figure}[htbp]
  \centering
  \includegraphics[width=\defaultfig,keepaspectratio,clip]{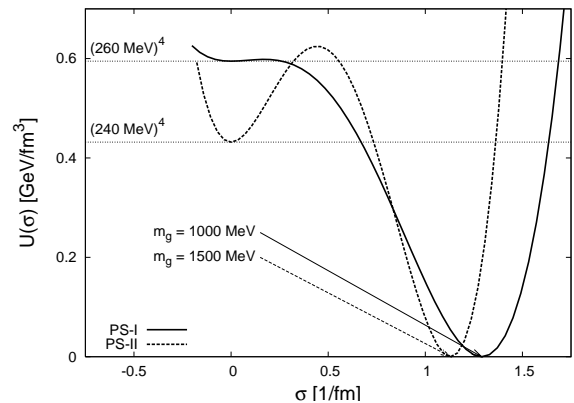}
  \caption{The scalar potential $U(\sigma)$ for the two different
  parameter sets PS1 and PS2 from \cite{Martens:2004ad} and stated
  explicitly in tab.~\ref{tab:params} further down.}
  \label{fig:ups12}
\end{figure}

The confinement field  is  exposed to a quartic self interaction
\begin{equation}
  \label{eq:uscalar}
  U(\sigma) = B + a\sigma^2 + b\sigma^3 + c\sigma^4.
\end{equation}
Two specific  forms
with  different parameters $B,a,b,c$ are shown in
fig.~\ref{fig:ups12}. The generic form of $U$ develops two (quasi-)
stable points, which separates the two distinct phases of the model. 
The first defines the
vacuum expectation value of the scalar field $\sigma =
\sigma_\text{vac}$ and is 
associated to the energy density of the confined phase with
$U(\sigma) = 0$. The second at $\sigma=0$ is associated to the
deconfined phase with an energy density $U(0) = B$. We refer to the
former phase as the non-perturbative vacuum and to the latter as the
perturbative vacuum. Given the generic form of the potential $U$ and
the  vacuum value $\sigma_\text{vac}$, there
are only two other independent parameters describing $U$. We choose
the perturbative value $B = U(0)$ and the curvature $m_g =
U''(\sigma)|_{\sigma_\text{vac}}$  of the potential at
$\sigma_\text{vac}$, where the primes denote  derivatives
with respect to $\sigma$. $m_g$ behaves as the mass of the confinement
field $\sigma$ and can be interpreted as the mass of a glueball, as
all non-perturbative gluonic effects are collected in the scalar field
of our model. Another possible interpretation is to relate the mass of
the confinement field to the mass of the off-diagonal gluons generated
in the maximal Abelian gauge \cite{Ezawa:1982bf,Amemiya:1998jz}. 
The parameters $a,b,c$ appearing in
Eq.~\eqref{eq:uscalar} can be expressed as
\begin{subequations}
  \label{eq:Uparameter}
  \begin{eqnarray}
    \label{subeq:Uapar}
    a &=& \frac{1}{2} \frac{m_g^2\sigma_\text{vac}^2 -
      12B}{\sigma_\text{vac}^2} \quad , \\
    \label{subeq:Ubpar}
    b &=& - \;\frac{m_g^2\sigma_\text{vac}^2 -
      8B}{\sigma_\text{vac}^3} \quad , \\    
    \label{subeq:Ucpar} 
    c &=& \frac{1}{2} \frac{m_g^2\sigma_\text{vac}^2 -
      6B}{\sigma_\text{vac}^4} \quad ,
  \end{eqnarray}
\end{subequations}
with the additional constraint, that $a\ge 0$ to ensure  that there is
no relative maximum at $\sigma = 0$.

The perturbative and the non-perturbative phases differ not only with
respect to the corresponding  energy densities $U(0)=B$ and
$U(\sigma_\text{vac})=0$, respectively, but also in their dielectric
behavior. In the former, the dielectric constant $\kappa(\sigma=0)=1$
allows for freely propagating fields, whereas in the latter
$\kappa(\sigma_\text{vac}) = \kappa_\text{vac} \ll 1$ and the electric
fields are suppressed. $\kappa_\text{vac} = 0$ would lead to perfect
screening of the color fields and the non-zero but small value of
$\kappa_\text{vac}$ is introduced for numerical reasons
\cite{Martens:2004ad}. The dielectric function $\kappa(\sigma)$ is
designed to interpolate smoothly between the two values and we choose
the following parameterization:
\begin{equation}
  \label{eq:kappa}
  \kappa(s) = \left\{
    \begin{array}{r@{\quad,\quad}l}
      1 + k_3 s^3 + k_4  s^4 + k_5 s^5 
      & 0 \le s \le 1\\
      1 & s < 0\\
      \kappa_\text{vac} & s > 1
    \end{array}\right. \quad ,
\end{equation}
with $s = \sigma/\sigma_\text{vac}$ and with coefficients
\begin{eqnarray}
  \label{eq:kappa_koeff}
  \left.
    \begin{array}{r@{\; =\;}l}
      k_3 & {\displaystyle \frac{1}{2} \left(29\kappa_\text{vac}- 20\right)}\\
      k_4 & {\displaystyle \phantom{\frac{1}{2}} 
        \left(15 - 23\kappa_\text{vac}\right)}\\
      k_5 & {\displaystyle \frac{1}{2} \left(19\kappa_\text{vac}- 12\right)}
    \end{array}
  \right\}
  \stackrel{\kappa_\text{vac}\rightarrow\, 0}{\longrightarrow}
  \left\{
    \begin{array}{r@{\; =\;}r}
      k_3 & {\displaystyle \vphantom{\frac{1}{2}} -10}\\
      k_4 & {\displaystyle \vphantom{\frac{1}{2}} 15}\\
      k_5 & {\displaystyle \vphantom{\frac{1}{2}} -6} \\
    \end{array}
  \right. ,
\end{eqnarray}
In the limit $\kappa_\text{vac} \rightarrow 0$ the non-perturbative
vacuum behaves as a perfect dielectric medium and all electric fields
are expelled out of regions where $\sigma=\sigma_\text{vac}$. Note,
however, that $\kappa_\text{vac} = 0$ is numerically not feasible. As
discussed in \cite{Martens:2004ad} there are different possibilities
to parameterize $\kappa(\sigma)$. With the polynomial form chosen
here, the results as shown below do not depend on the exact value of
$\kappa_\text{vac}$, once it is smaller than $\kappa_\text{vac} <
10^{-3}$. In this work we choose $\kappa_\text{vac} = 10^{-4}$.

The fields $\phi^a$ and $\sigma$ and the corresponding field energies
are calculated numerically according to Eqs.~\eqref{eq:static_eoms}
and \eqref{eq:energy_decomp} using the multi-grid FAS-algorithm given
in \cite{brandt82,nrc96,Martens:2004ad}.

The dependence of the numerical results on the model parameters was
studied in detail in \cite{Martens:2004ad}. Different sets of
parameters were given that reproduce  the Cornell
potential for a $q\bar{q}$-pair and simultaneously the profile of the
color electric string. The Cornell potential was found in heavy
quarkonium spectroscopy
\cite{Eichten:1975af,Quigg:1979vr,Eichten:1978tg,Eichten:1980ms} and
was reproduced on the lattice both in SU(2) \cite{Bali:1995de} and in
SU(3) \cite{Bali:1992ab} and is given by
\begin{equation}
  \label{eq:cornell}
  V_c(R) = 2 C_F E_0 - C_F \frac{\alpha}{R} 
  + \tau  R \quad .
\end{equation}
We focused on reproducing the string tension $\tau=980\,$MeV and the
Coulomb coefficient 
$\alpha=0.3$. Here $R$ is the distance between the quark and the
anti-quark and $C_F = 1/3$ is the corresponding color factor in the
Abelian approximation. The constant and finite term $E_0$ is due to
the non-zero width of the particles. The transverse profile was
calculated on 
the lattice \cite{Bali:1995de}, in the framework of the dual
color superconductor \cite{Baker:1991bc,Maedan:1990ju} and in the
Gaussian Stochastic Model
\cite{Kuzmenko:2000rt,Shoshi:2002rd}. The profile is
rather well described by a Gaussian-like parameterization 
\begin{equation}
  \label{eq:gaussian}
  f_g(\rho) = N_g \exp[-\ln 2\,(\rho/\rho_g)^n] \quad ,   
\end{equation}
with a half width $\rho_g\approx
0.3\,$fm and a steepness parameter ranging from $n=2.3$ to $n=3.2$. In
tab.~\ref{tab:params} we give two of the parameter sets used  in
\cite{Martens:2004ad}, together with the given key quantities of the
$q\bar{q}$-string.

\begin{table*}[htbp]
  \centering
  \begin{tabular}{c||c|c|c|c|c||c|c|c|c}
    No. & $B^{1/4}$ [MeV] & $m_g$ [MeV] & $\sigma_\text{vac}
    \,\left[\text{fm}^{-1}\right]$ 
    & $g_s$ & $\kappa_\text{vac}$
    & $\tau \left[\frac{\text{MeV}}{\text{fm}}\right]$ & $C_F\alpha$ &
    $\rho_g$ [fm] & $n$\\\hline  
    I   & 260  & 1000  & 1.29  & 2.0 & $10^{-4}$
    & 980 & 0.18 & 0.33 & 2.3  \\
    II  & 240 & 1500  & 1.13  & 1.8 & $10^{-4}$
    & 980 & 0.12 & 0.34 & 3.1  \\
  \end{tabular}
  \caption{In the first 5 columns we show the CDM parameter sets used
    in the description of $q\bar{q}$-strings and $qqq$-baryons, in the
    last 4 columns we list the 
    resulting values for the string tension $\tau$, the
    Coulomb-parameter $\alpha$ and the  shape parameters of the
    profile, i.e.~the width $\rho_g$ and the steepness parameter $n$
    as explained in the text.}
  \label{tab:params}
\end{table*}

All basic quantities of the Cornell potential and the profile  of the
$q\bar{q}$-string agree either with lattice results or with results
obtained in heavy meson spectroscopy, except for the Coulomb constant
$\alpha$ which is somewhat small in our results.

\section{Long range behavior of $q\bar{q}$-strings}
\label{sec:long-range}

\begin{figure}[htbp]
  \centering
  \includegraphics[width=\defaultfig,keepaspectratio,clip]%
  {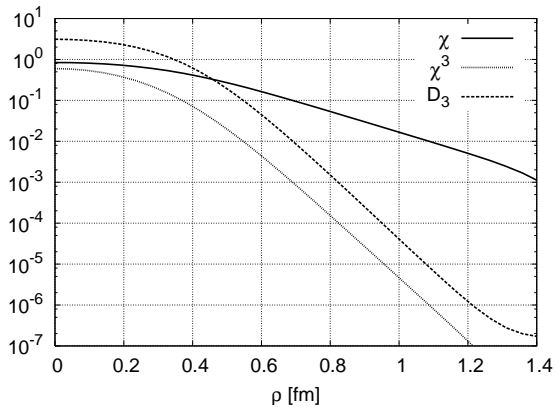} 
  \caption{The logarithmic profile of the string. Shown is the profile
    of the confinement field $\chi=\sigma_\text{vac}-\sigma$ (solid
    line, cf.~Eq.~\eqref{eq:chi-definition}) and of the 3-component of the
    electric field $\vec{D}_3$. The slope of the $\vec{D}_3$-field
    is much  steeper and is equal to that of the third power $\chi^3$
    of the confinement field (dotted line).}
  \label{fig:string_profile}
\end{figure}

In \cite{Martens:2004ad} we performed a detailed analysis of
$q\bar{q}$-strings with the main focus on the bulk properties of the
strings such as the string tension $\tau$ and the width of the strings
$\rho_g$. The profile of the energy density, which we determine
numerically by solving the Eqs.~\eqref{eq:static_eoms}, 
was well described by a generalized Gaussian form as in
Eq.~\eqref{eq:gaussian}. On a linear scale the deviations of the
numerical results to the Gaussian fit were hardly seen. However, this
parameterization does not reproduce the long range behavior of the
string fields far away from the string axis. Instead both the electric
fields $\vec{D}^a$ and the confinement field $\sigma$ follow an
exponential as can be seen in fig.~\ref{fig:string_profile}. It should
be noted that the exponential tail of the string fields does not
influence the results obtained in \cite{Martens:2004ad}. 
This exponential behavior can be explained quite naturally, if one
assumes, that the electric fields die out on a much shorter
characteristic length scale than the confinement field. With this
assumption the source term in the right hand side of
Eq.~\eqref{eq:static_sigma} vanishes. For small deviations
\begin{equation}
  \label{eq:chi-definition}
  \chi = \sigma_\text{vac} - \sigma
\end{equation}
of the confinement field
from its vacuum value we can make a Taylor expansion of $U(\sigma)$
around $\sigma_\text{vac}$ leading to $U(\sigma) \approx \half m_g^2
(\sigma - \sigma_\text{vac})^2$. 

\begin{figure}
  \centering
  \includegraphics[width=\defaultfig,keepaspectratio,clip]%
  {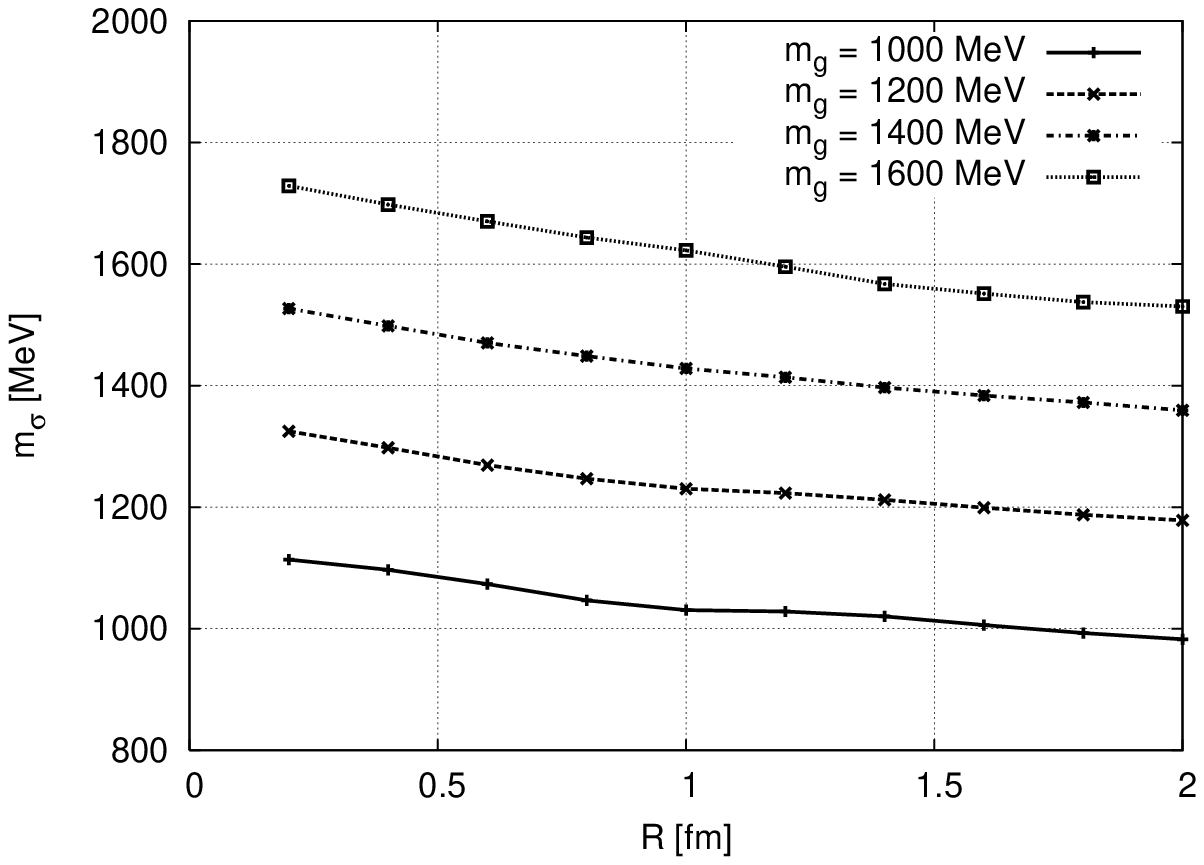}
  \includegraphics[width=\defaultfig,keepaspectratio,clip]%
  {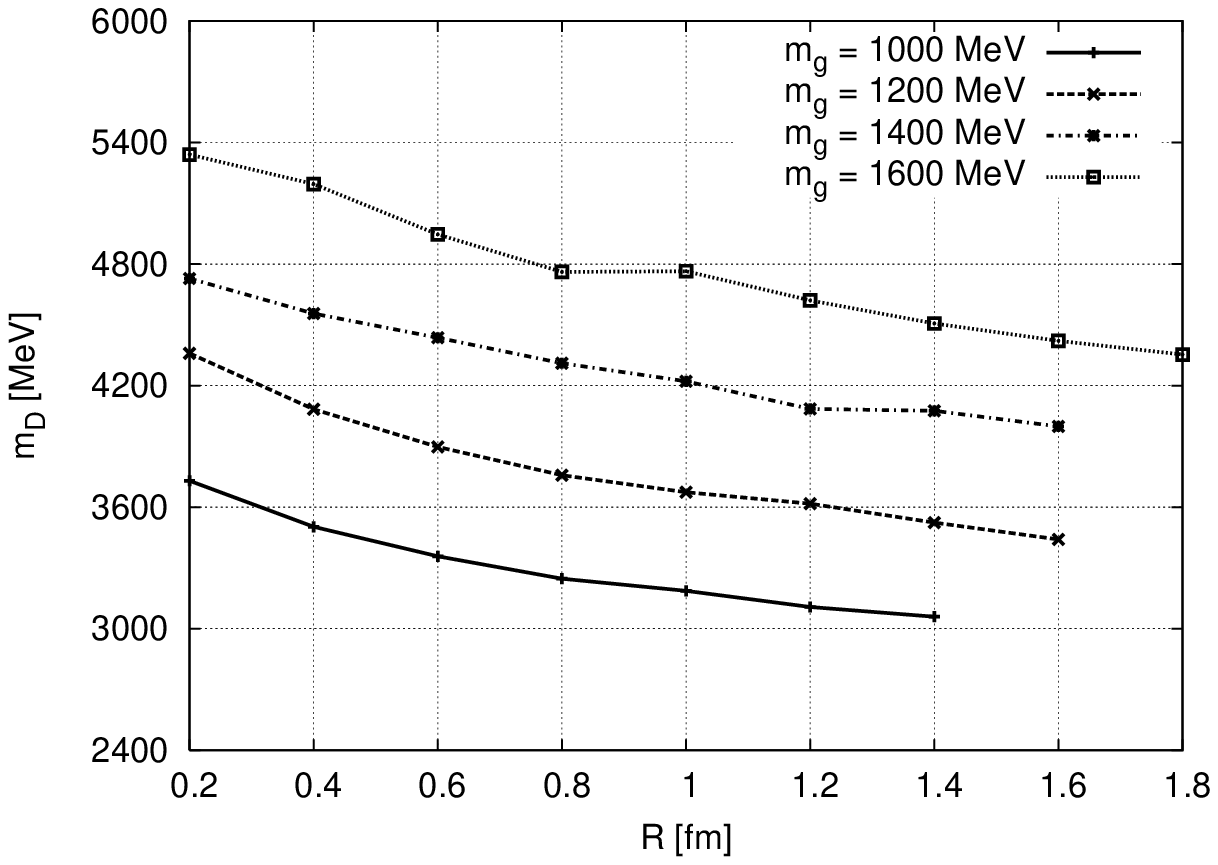}
  \caption{The fitted values of $m_\sigma$ (upper panel) and $m_D$
    (lower panel) for different model parameters $m_g$ and for
    different string lengths $R$.} 
  \label{fig:mass-fit}
\end{figure}

For strings with very large $q\bar{q}$-separations $R$ we can recast
Eq.~(\ref{eq:static_sigma}) into
\begin{equation}
  \label{eq:sigma_cylinder}
  \frac{1}{\rho} \frac{\partial}{\partial\rho}
  \left(
    \rho\frac{\partial\chi}{\partial\rho}
  \right)
  - m_g^2 \chi = 0 \quad ,
\end{equation}
where $\rho$ is the coordinate transverse to the string axis and where
we have used cylindrical symmetry. The regular solution of this
equation is
\begin{equation}
  \label{eq:sigma-sol}
  \chi(\rho) = \chi_0 K_0(m_g\rho) \approx 
  \chi_0 {\textstyle \sqrt{\frac{\pi}{2m_g\rho}}} \; e^{-m_g\rho}
  \quad ,
\end{equation}
where $K_0(x)$ is the modified Bessel function of the second kind and the
second relation holds for large $\rho$ and $\chi_0$ is some
constant. The parameter $m_g$ is 
therefore directly connected to the screening mass $m_\sigma = m_g$ of
the confinement field or to its screening length $\xi = 1/m_g$. The electric
field is  $\vec{D}^a = \kappa(\sigma) \vec{E}^a$. In the chosen
parameterization the function $\kappa(\sigma)$ itself and its first two
derivatives at $\sigma=\sigma_\text{vac}$ are proportional 
to the numerically small number $\kappa_\text{vac}$. For small
deviations $\chi=\sigma_\text{vac}-\sigma$ we expect 
\begin{subequations}
  \label{eq:Dfield-sol}
  \begin{eqnarray}
    \kappa(\sigma) &=& \kappa_0 (\chi/\chi_0)^3 + {\cal
    O}(\kappa_\text{vac})\\
    D^a(\rho) &=& \kappa_0 \, (\chi/\chi_0)^3 \, E \nonumber\\
    \label{subeq:Dfield-paramtion}
     &=& \kappa_0 {\textstyle \sqrt{\frac{\pi}{2m_g\rho}}^{\,3}} \;  
    e^{- m_D \rho} \; E \\
    m_D &=& 3 m_g
  \end{eqnarray}  
\end{subequations}
where $\kappa_0$ is a proportionality constant and $E$ is the electric
field which varies only slowly with $\rho$. The screening
mass of the electric displacement is therefore $m_D = 3m_\sigma = 3
m_g$ and the field is screened on a characteristic length scale
$\lambda = \frac{1}{3}\xi$. This justifies a posteriori to neglect in
Eq.~\eqref{eq:sigma_cylinder} the source term on the right-hand side of
Eq.~\eqref{eq:static_sigma}. We have added in
fig.~\ref{fig:string_profile} the third power $\chi^3$ of $\chi$ 
(dotted line) which shows the same slope as $D^{(a=3)}$ (dashed line). 
To check this result numerically we have fitted the above analytical
solutions in Eqs.~\eqref{eq:sigma-sol} and
\eqref{subeq:Dfield-paramtion} to both the confinement field $\chi$
and the electric displacement $D^a$ obtained in our numerical
calculations leaving the parameters $m_\sigma$ and $m_D$ 
as fit parameters. In fig.~\ref{fig:mass-fit} we show the results for
different model parameters $m_g$ and for different string lengths
$R$. The dependence of $m_\sigma$ and $m_D$ on $m_g$ for fixed $R$ is
obvious. For fixed $m_g$ the fitted values approach roughly the
expected values $m_\sigma = m_g$ and $m_D = 3m_g$, respectively, for
growing string lengths $R$. Actually the fitted values undershoot the
theoretical values consistently, which might be due to the final
extent of the numerical box and the numerical boundary conditions
$\chi = 0$ on the box boundary.

Both the confinement field $\sigma$ and the electric displacement $D^a$
follow an exponential far away from the string axis. They vanish on 
characteristic screening lengths with $\lambda = (3m_g)^{-1} < \xi =
m_g^{-1}$. This is somewhat similar to a dual type I color
superconductor, where $\lambda$ and $\xi$ describe the penetration
depth of the  color field and the correlation length of the Higgs
field. In \cite{Matsubara:1994nq} and \cite{Bali:1998cp} a lattice
analysis of the string fields in Abelian projection was made and a
similar result $\lambda \apprle \xi$ was obtained. However, it should
be noted, that $m_D = 3m_g$ depends on the specific behavior of
$\kappa(\sigma)$ at $\sigma=\sigma_\text{vac}$ and might be modeled
differently. In contrast $m_\sigma=m_g$ follows from the e.o.m. as
long as the $D^a$-field can be neglected in
Eq.~\eqref{eq:static_sigma}. 

\section{String interactions}
\label{sec:stringint}

In the CDM strings or flux tubes develop between two oppositely
charged quarks, or more generally, between a number of particles with
vanishing total net color charge. In the case of four particles, there
must be two quarks and two anti-quarks to ensure color neutrality. For
the discussion of string interactions the network of flux tubes does
depend strongly on the configuration in coordinate space but
also on the configuration in color space. In the latter there are two
distinct possibilities. First, all quarks can be of the same color
$|c\rangle$ and the corresponding anti-quarks have the anti-color
$|\bar{c}\rangle$, and thus the color content is
($cc\bar{c}\bar{c}$). Second, there is one string with color content
$c\bar{c}$ and another one with $c'\bar{c}'$ with $c \neq c'$, so that the
color content is ($cc'\bar{c}\bar{c}'$). This latter configuration is
not possible in SU(2), where the two members of the fundamental
doublet are simultaneously the anti-particles to each other. In this
work we concentrate on the former color configuration to compare our
results to those obtained in SU(2) lattice calculations
\cite{Green:1993ag,Green:1994hk,Pennanen:1998nu} and in the model of
the dual color superconductor \cite{Kodama:1997zc}. 

\begin{figure}[htbp]
  \centering
  \includegraphics[width=\defaultfig,keepaspectratio,clip]{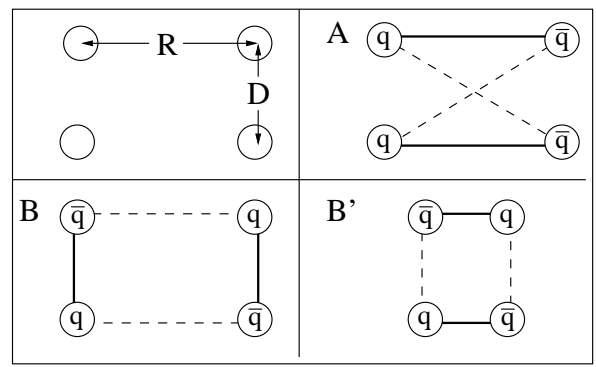}
  \caption{Upper panel: Four particles placed on the corners of a
    rectangle with length $R$ and width $D$ (left). Two parallel
    strings of length $R$  and distance $D$ ($A$). Lower panel: Two
    anti-parallel strings with $R>D$ ($B$) and $R<D$ ($B'$). Solid lines
    stand for the ground state of the two possible flux tubes, dashed
    lines for a possible excited state.} 
  \label{fig:stringAB}
\end{figure}

In coordinate space the orientations of the flux tubes depend on the
actual distributions of colors and anti-colors. We first discuss the
simple case of four particles with colors $cc\bar{c}\bar{c}$ placed on
the corners of a rectangle with length $R$ and width $D$ as shown in
the upper left part of fig.~\ref{fig:stringAB}. If all quarks $q$ are
lying on the left side and the anti-quarks $\bar{q}$ on the right side
the ground state of flux tubes will be like in configuration $A$ in the
same figure independent of $R$ and $D$. In this configuration we have
two strings of length $R$ with the electric flux pointing in the same
direction. In varying $D$ we can examine the interaction energy of two
strings of given length $R$ and thus obtain the static string
potential for this particular distribution. In \cite{Loh:1997sk} the
potential between two $q\bar{q}$-strings was deduced with a dynamical
setup, albeit with the oversimplifying assumption, that the electric 
fields of two strings add coherently, thus neglecting the intrinsic
four-particle interactions. 

If in contrast the positions of a quark
and an anti-quark are interchanged, there exist two different
possibilities. As the energy of a single $q\bar{q}$-string is a
monotonically rising function of separation $R$
\cite{Quigg:1979vr,Bali:1997am,Martens:2004ad}, the lowest four-particle
energy is obtained by minimizing the total string length. Consequently if
$R>D$, the flux tubes point upside down as in configuration $B$ in
fig.~\ref{fig:stringAB}, whereas, if $R<D$, the flux tubes point from
left to right as in configuration $B'$. In the CDM the ground state of
the flux tubes is found within the model itself by solving the
e.o.m.~\eqref{eq:static_eoms}, without further input. If
one increases $D$ with fixed $R$, the strings will flip from
configuration $B$ to $B'$ when $D=R$, and one can study the string
flip interaction energy. In addition, a dynamical string flip can lead
to the dissociation of a $J/\psi$ and thus to $J/\psi$ suppression in
relativistic heavy ion collisions \cite{Loh:1997cr}.

We define the interaction between two strings as the difference
between the 4-particle energy $E_4$ and the sum of the energies
$E_{s_1} + E_{s_2}$ of two independent strings $s_1$ and $s_2$
with minimal energy. In the case of four particles
$q_1q_2\bar{q}_1\bar{q}_2$ we calculate the sum of the energies
$E_{q_1\bar{q}_1}$  and $E_{q_2\bar{q}_2}$ of the strings
($q_1\bar{q}_1$) and ($q_2\bar{q}_2$)  separately and compare it to
the sum of the energies $E_{q_1\bar{q}_2}$  and $E_{q_2\bar{q}_1}$ of 
the competing pairs ($q_1\bar{q}_2$) and ($q_2\bar{q}_1$). In the
absence of any interactions, the lower
value defines the ground state energy of the strings $s_1$ and $s_2$
(solid lines in fig.~\ref{fig:stringAB}), the higher an excited
state (dashed lines in fig.~\ref{fig:stringAB}). As the energy of a
$q\bar{q}$-string rises monotonously, we do not have to solve the
equation of motion for both configurations but only for that with the
minimal total string length. The interaction potential
$V_4$ is then given by  
\begin{subequations}  
  \label{eq:potential}
  \begin{eqnarray}
    V_4 &=& E_4 - (E_{s_1} + E_{s_2}) \quad ,\\
    E_{s_1} + E_{s_2} &=& \min\left\{
      \begin{array}{ll}
        E_{q_1\bar{q}_1} + E_{q_2\bar{q}_2}\\
        E_{q_1\bar{q}_2} + E_{q_2\bar{q}_1}  \quad .
      \end{array}
    \right.
  \end{eqnarray}
\end{subequations}

To illustrate, how the two strings of the ground state influence each
other we can define the spatial distribution of the interaction energy
$v_4$ as:
\begin{equation}
  \label{eq:inter-density}
  v_4(\vec{r}) = \varepsilon_4(\vec{r}) - 
  (\varepsilon_{s_1}(\vec{r}) + \varepsilon_{s_2}(\vec{r})) \quad ,
\end{equation}
where the energy densities $\varepsilon$ are given as the integrands
of Eqs.~\eqref{subeq:energy_el}-~\eqref{subeq:energy_sur}. One comment
about the special configuration where the particles are located on the
corners of a square is needed. If the two strings are of type $B$/$B'$ in
fig.~\ref{fig:stringAB}, both configurations are degenerate for $R=D$,
that is the sums of energies $E_{q_1\bar{q}_1} + E_{q_2\bar{q}_2}$ and
$E_{q_1\bar{q}_2} + E_{q_2\bar{q}_1}$ are equal. In that case we
calculate the interaction energy density as
\begin{eqnarray}
  \label{eq:inter-density-square}
  v_4(\vec{r}) &=& \varepsilon_4(\vec{r}) - 
  \frac{1}{2}\big[(\varepsilon_{q_1\bar{q}_1}(\vec{r}) 
  + \varepsilon_{q_2\bar{q}_2}(\vec{r}))  \nonumber\\
  && \hphantom{\varepsilon_4(\vec{r}) - }
  + (\varepsilon_{q_1\bar{q}_2}(\vec{r}) 
  + \varepsilon_{q_2\bar{q}_1}(\vec{r}))\big] \quad ,
\end{eqnarray}
i.e.~we compare the four-particle density with an incoherent
superposition of the 2-string configurations $B$ and $B'$. Of course this
matters only in the description of the energy \emph{density} and not
of the potential energy $V_4$. 

Given the exponential behavior of the strings far away from the string
axis (cf. Eqs.~\eqref{eq:sigma-sol} and \eqref{eq:Dfield-sol}),
one might estimate the interaction between two separated strings in
the following way: Assume both strings point along the x-axis with
their axes shifted by $\pm D/2$ away from the string axis. For
sufficiently large $D$, the approximations made in section
\ref{sec:long-range} are valid and the equation for the confinement
field is linearized as in Eq.~\eqref{eq:sigma_cylinder}. The
confinement field of the two string configurations at $|y| \ll D$
then simply is a linear superposition of two single strings, i.e.

\begin{subequations}
  \label{eq:two-string-sigma}
  \begin{eqnarray}
    \chi(y) &=& \chi_1(y) + \chi_2(y) \quad \text{with} \\
    \chi_{1/2} &=& \frac{\chi_0}{\sqrt{m_g(\frac{D}{2}\pm y)}} 
    e^{\mp m_g(y \pm D/2)} \quad ,
  \end{eqnarray}  
\end{subequations}
with $\chi_0$ some constant. As the screening length $\lambda$ of the
electric field is substantially smaller than that of the confinement
field, $\lambda < \xi$, the electric fields do not contribute much to
the total energy density $\varepsilon$, which consequently is
dominated by the scalar potential $U(\chi) \approx \half m_g^2
\chi^2$. The distribution of the interaction energy $v_4$ (see
Eq.~\eqref{eq:inter-density}) can be expressed with 
Eq.~\eqref{eq:two-string-sigma} as 
\begin{eqnarray}
  v_4(y) &=& \half m_g^2
  \left(
    \chi^2(y) - \chi_1^2(y) - \chi_2^2(y)
  \right) \nonumber\\
  &\approx& 2 m_g^2 \chi_0^2 \; \frac{e^{-m_g D}}{m_g D} \nonumber\\
  \label{eq:string-interaction}
  &\approx& v_0 \frac{e^{-m_g D}}{m_g D} \quad ,
\end{eqnarray}
with $v_0$ some constant and where we have used $|y| \ll D$. The
strings at a distance $D$ therefore interact with a Yukawa potential,
which will be verified numerically later on.

\subsection{Electric field $\vec{D}$}
\label{sec:fields}

\begin{figure}[htbp]
  \centering
  \includegraphics[width=\narrowfig,keepaspectratio,clip]%
  {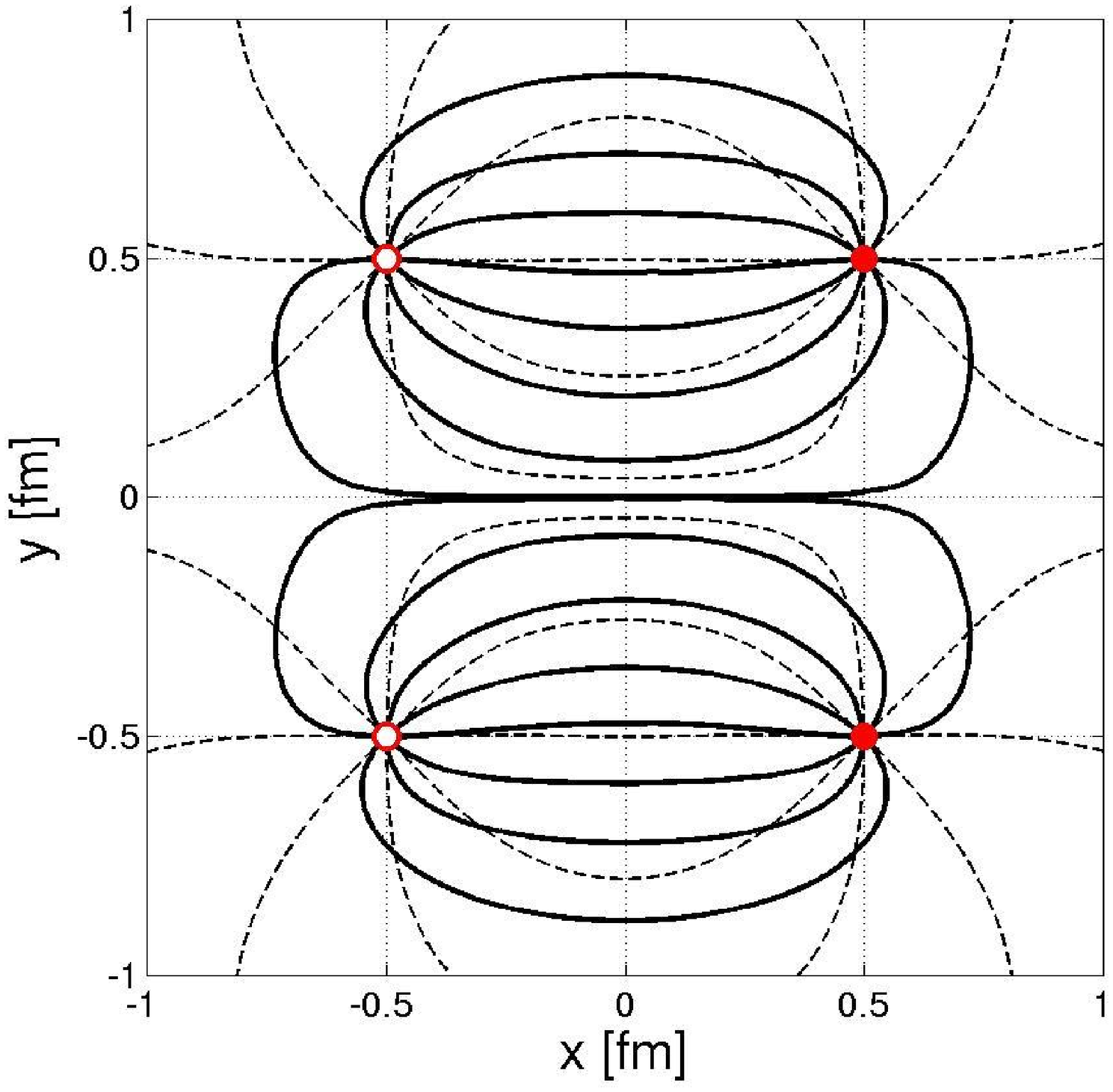}
  \includegraphics[width=\narrowfig,keepaspectratio,clip]%
  {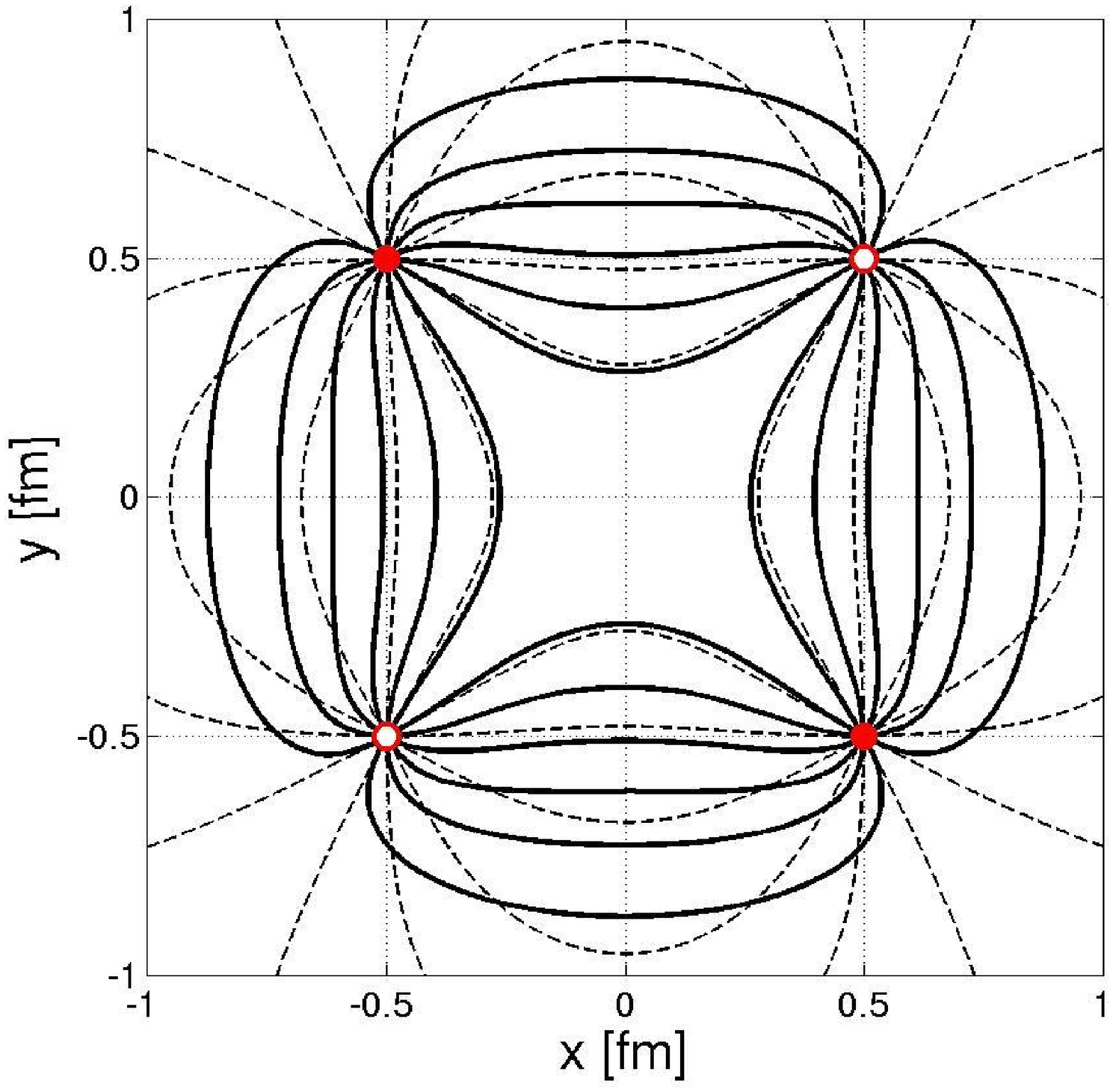}
  \caption{The electric displacement $\vec{D}$ for
    $\kappa=\kappa(\sigma)$ (solid lines) and for $\kappa = 1$ (dashed
    lines). The strings have a length $R=1\,\text{fm}$ and their
    distance is $D=1\,\text{fm}$. The upper and the lower plot show a
    parallel and an anti-parallel orientation, respectively. The
    confinement mechanism of the CDM pushes the field lines into well
    defined flux tubes. For the anti-parallel case the electric flux is
    distributed symmetrical into types $B$ and $B'$.} 
  \label{fig:dfield-kappa}
\end{figure}

We start our numerical analysis by discussing the electric field lines
of two 4-particle configurations with square symmetry in
fig.~\ref{fig:dfield-kappa}, where the particles are located at the
corners of a square ($R=D=1\,$fm). In this and the following figures 
we have a planar particle configuration, although the calculations are
done in three dimension. The figures show a cut along the plane $z=0$
of the particles. If not stated otherwise, the calculations for the
next figures were done using parameter set PS1 from
table~\ref{tab:params}. In the  
upper panel, the strings are parallel to each other (type $A$ of
fig.~\ref{fig:stringAB}) and in the lower panel, the strings are
anti-parallel to each other (type $B$/$B'$ in fig.~\ref{fig:stringAB}). In
both figures the field lines are shown both for the case
$\kappa(\vec{r}) = \text{const}=1$, i.e.~for the color fields obeying
the ordinary electromagnetic 
Maxwell equations (dashed lines), and for $\kappa[\sigma(\vec{r})]$ as
calculated from Eqs.~\eqref{eq:static_eoms} (solid lines). All
field lines are chosen to start on a circle around each quark (filled
dots) with equal angular separation and consequently end on the
anti-quarks (open dots). For both orientations the field lines in the
CDM calculations are compressed into well defined flux tubes
stretching from quarks to anti-quarks, as opposed to the $\kappa=1$
case, where the field lines extend in all directions. The flux tubes
of the anti-parallel configuration (lower panel) split symmetrically
into 
two parts connecting each quark with both anti-quarks. Therefore we
have a superposition of the two 2-string configurations $B$ and $B'$ of
fig.~\ref{fig:stringAB}. Due to the
symmetry of the configuration, there is no electric field at all in
the center at $\vec{r}=0$.

\subsection{Energy distribution}
\label{sec:distribution}

\begin{figure}[htbp]
  \centering
  \includegraphics[width=\defaultfig,keepaspectratio,clip]{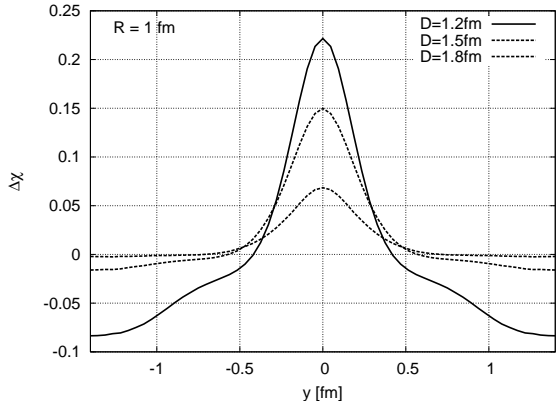}
  \caption{The difference $\Delta\chi = [\chi - (\chi_1 +
    \chi_2)]/\chi$ (cf. Eq.~\eqref{eq:two-string-sigma}) for a type A
    configuration of two parallel strings of length
    $R=1\,\text{fm}$. The plot shows a cut along the $y$-axis
    transverse to the string axes through the center of the
    configuration.}
  \label{fig:4q-sigma}
\end{figure}

To analyze the validity of our assumptions in
Eqs.~\eqref{eq:two-string-sigma} and \eqref{eq:string-interaction} we
show the relative difference $\Delta\chi = [\chi - (\chi_1 +
\chi_2)]/\chi$ (see 
Eq.~\eqref{eq:two-string-sigma}) for a configuration of two 1
fm long parallel strings in fig.~\ref{fig:4q-sigma}. As expected, the
difference approaches 0 for increasing $D$, i.e. the scalar field in
the center between two strings is a linear superposition of two
isolated strings for sufficiently large $D$. Note, that $\Delta\chi$
becomes concentrated at $|y| \ll D$ for large $D$, i.e. the fields of
the strings cancel each other at the centers of each string.  

\begin{figure*}[htbp]
  \centering
  \includegraphics[width=0.32\widefig,keepaspectratio,clip]%
  {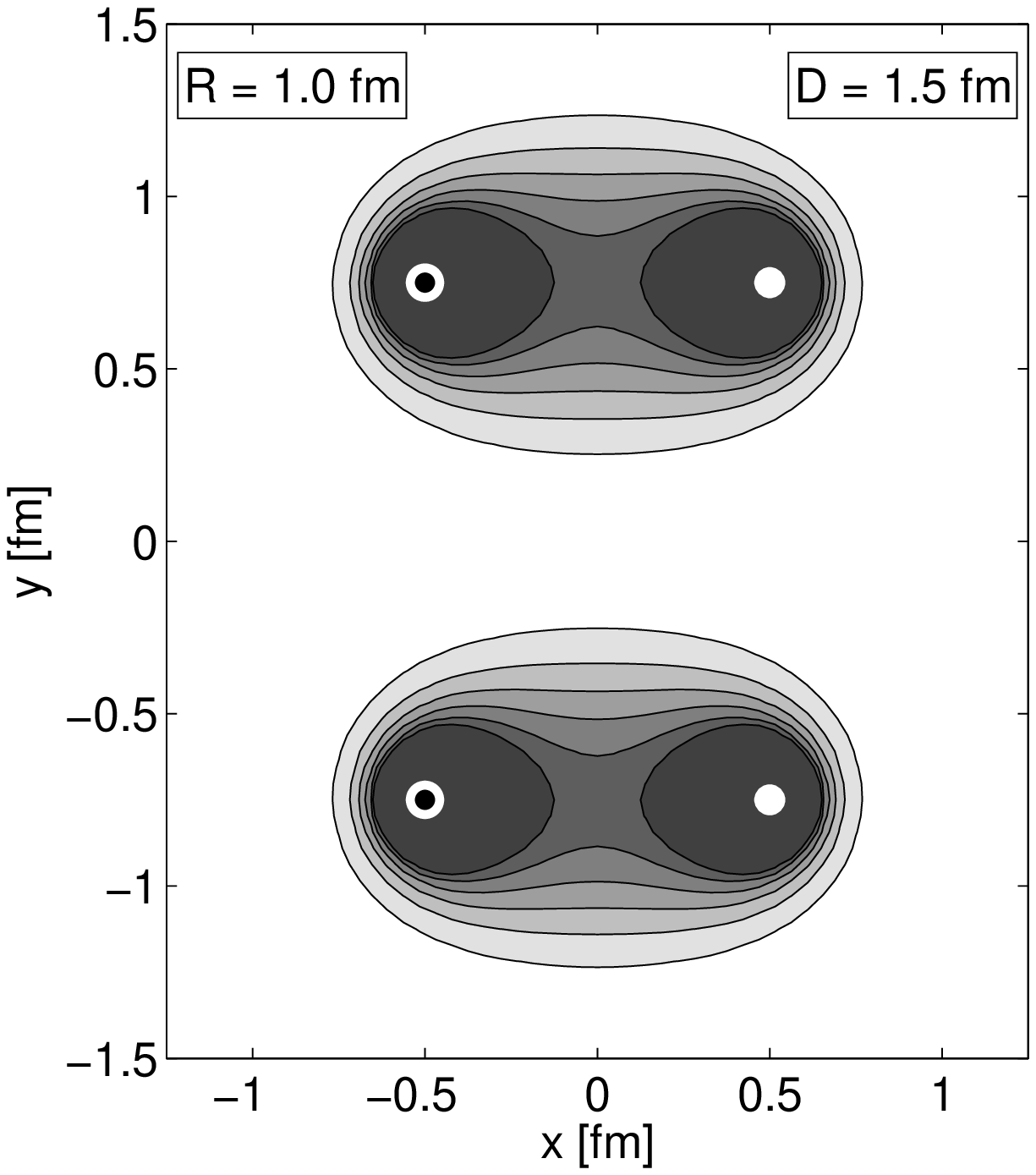}\hfill
  \includegraphics[width=0.32\widefig,keepaspectratio,clip]%
  {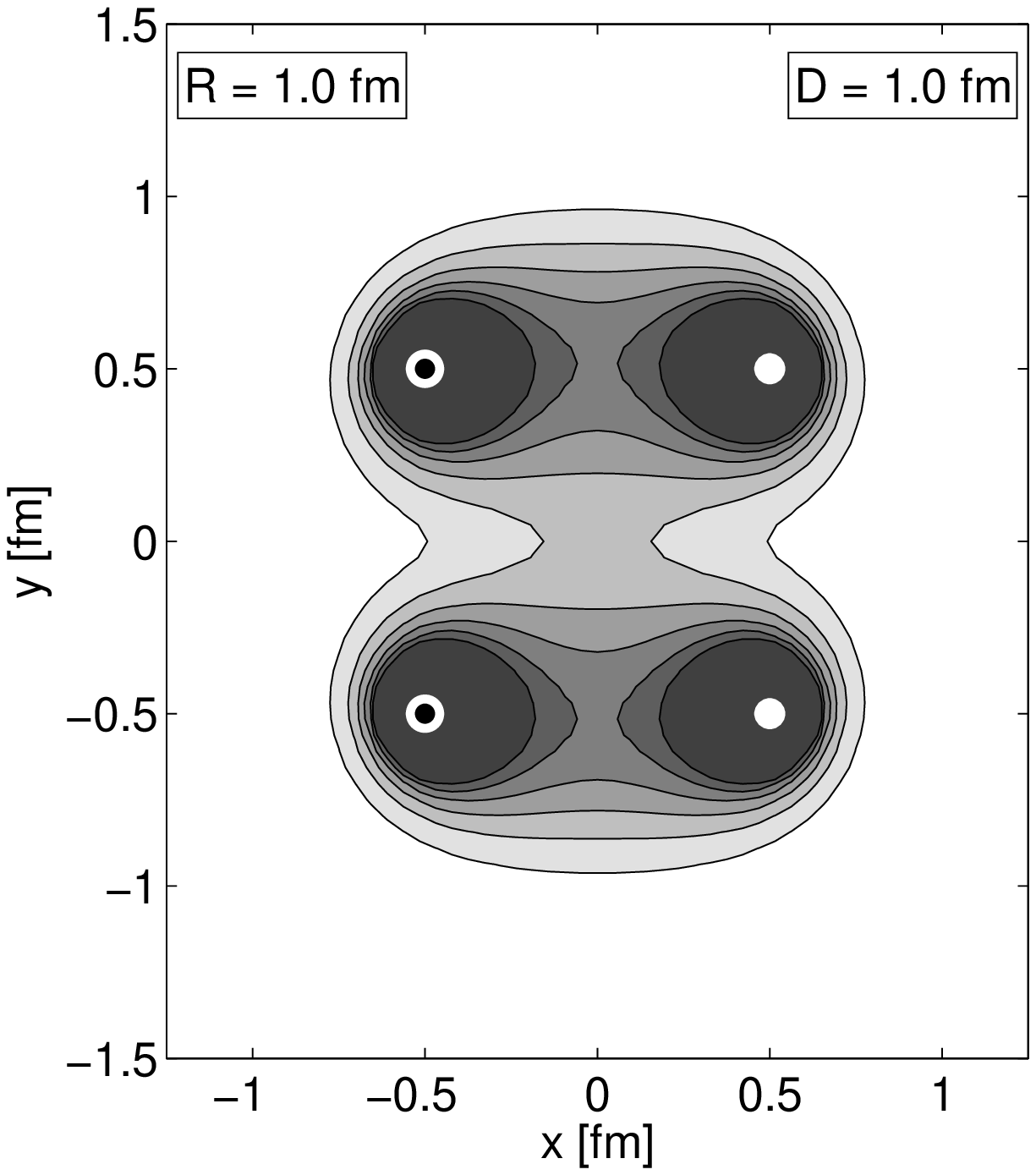}\hfill
  \includegraphics[width=0.32\widefig,keepaspectratio,clip]%
  {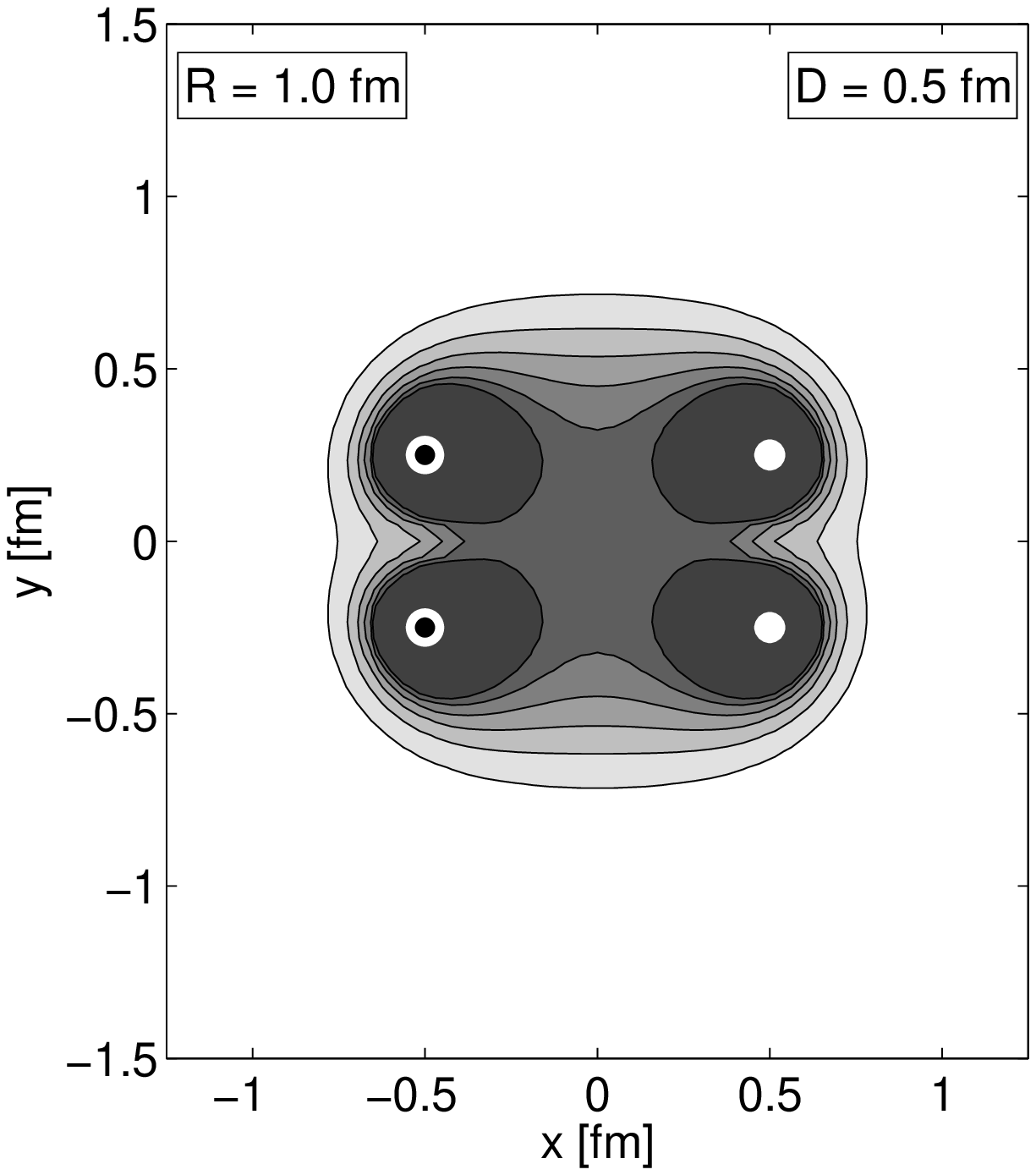}

  \includegraphics[width=0.32\widefig,keepaspectratio,clip]%
  {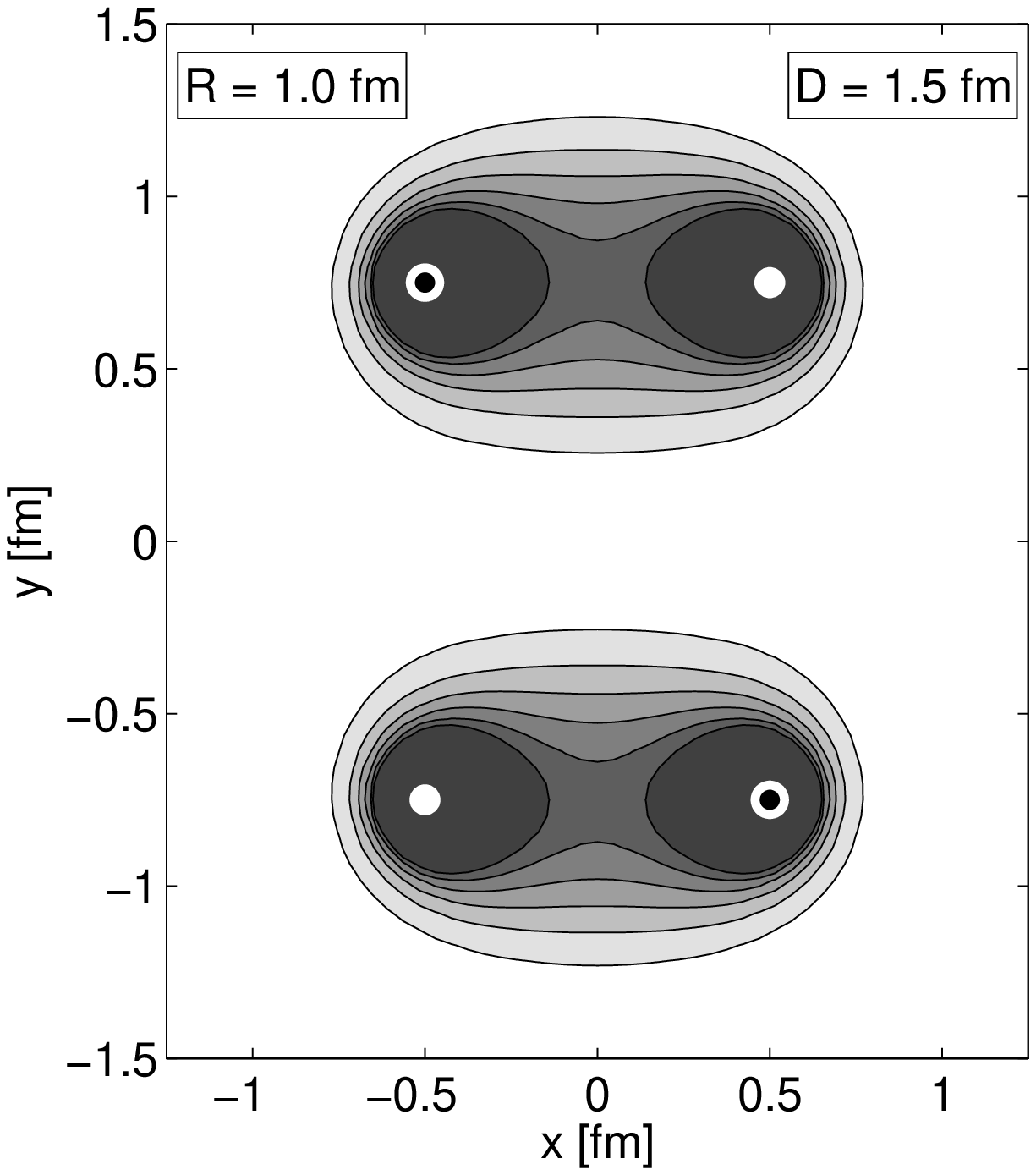}\hfill
  \includegraphics[width=0.32\widefig,keepaspectratio,clip]%
  {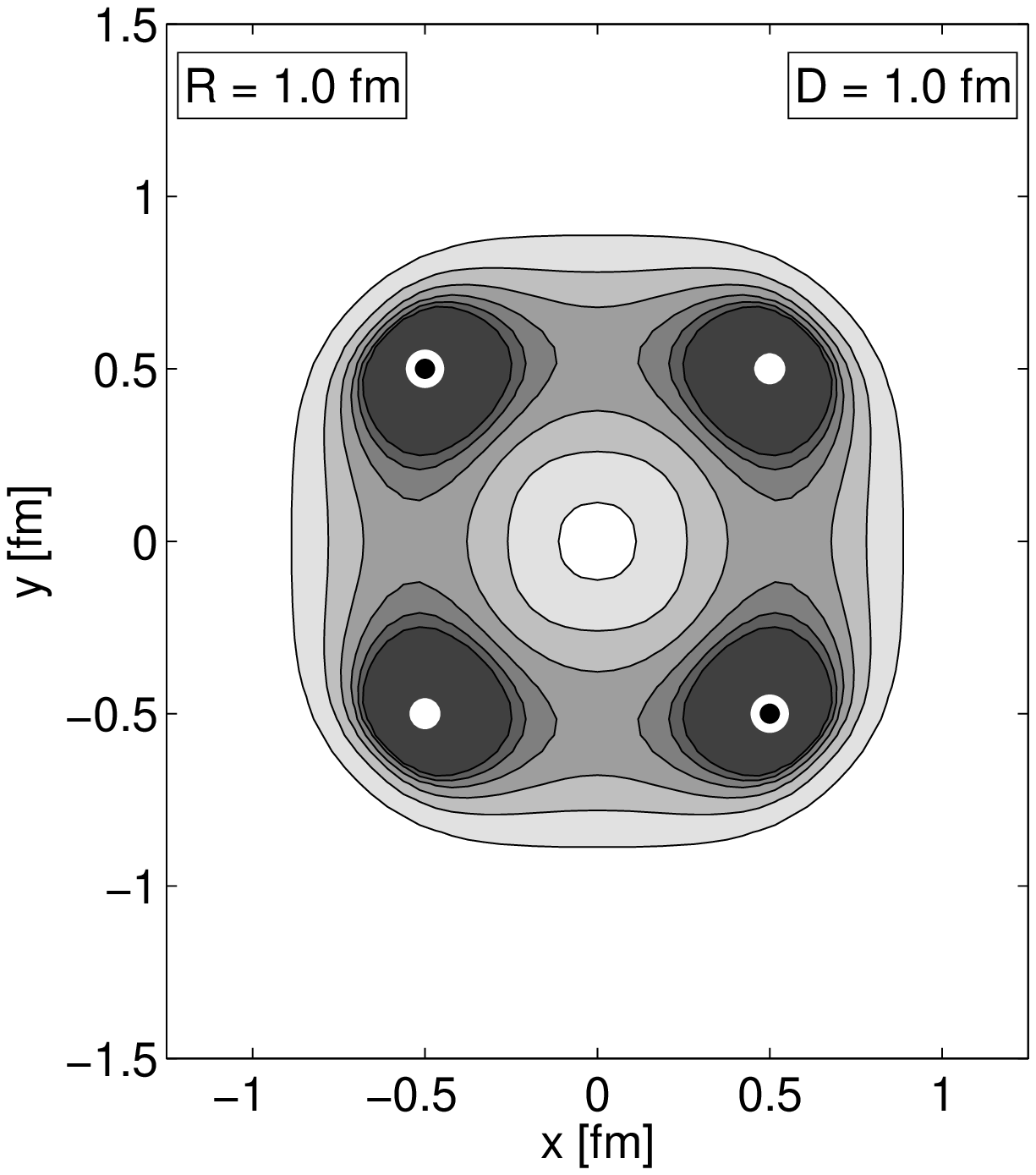}\hfill
  \includegraphics[width=0.32\widefig,keepaspectratio,clip]%
  {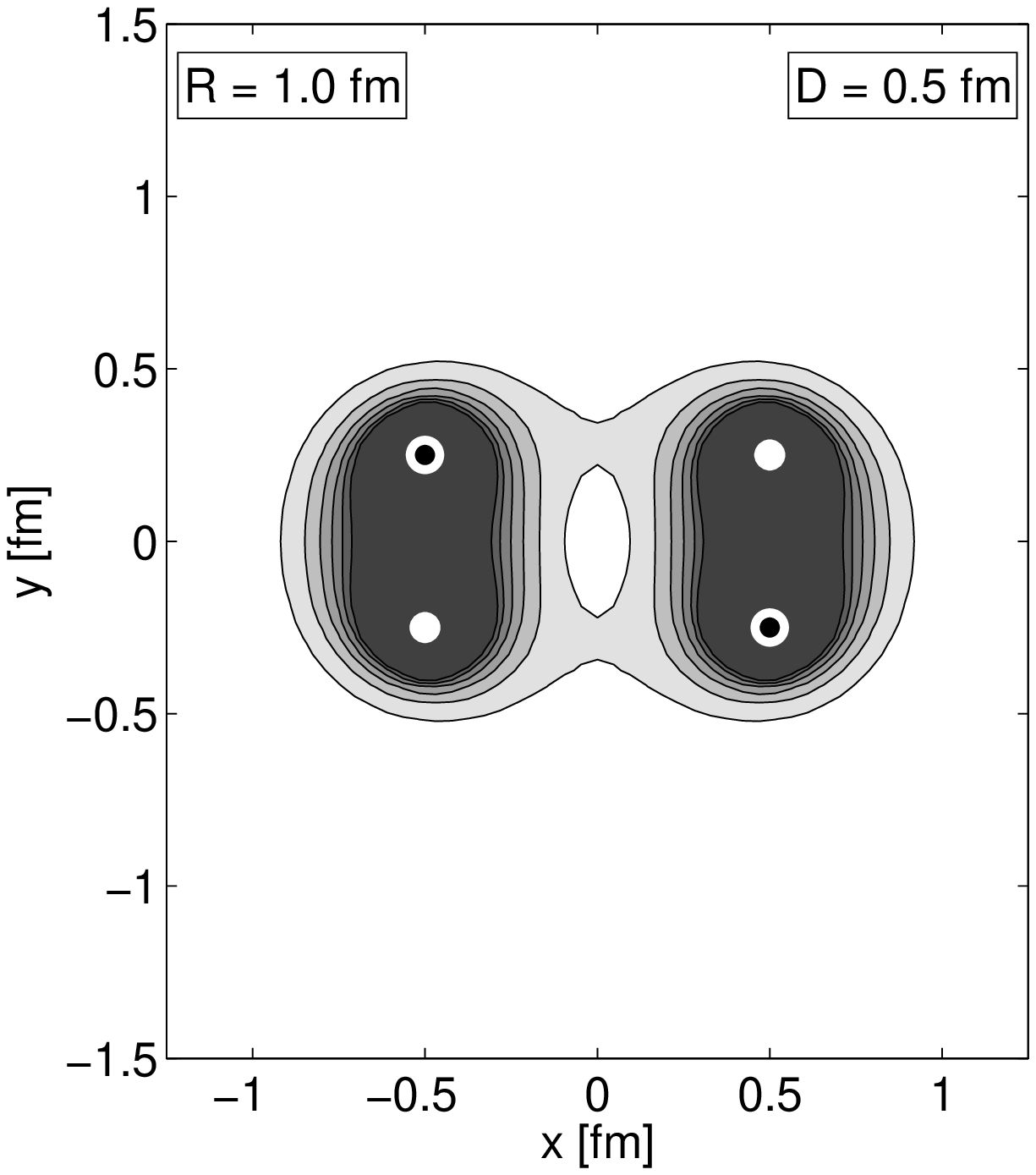}
  \caption{Energy density of two $R=1\,\text{fm}$ long
    $q\bar{q}$-strings for different distances $D$. 
    The orientation of the strings is parallel
    (upper panel) and anti-parallel (lower panel). Contour lines are
    chosen at the values $\varepsilon=(0.4\,\text{GeV/fm}^3 \ldots
    2.4\,\text{GeV/fm}^3)$ in steps of
    $\Delta\varepsilon=0.4\,\text{GeV/fm}^3$. The darker the area the
    larger is $\varepsilon$. Quarks are depicted as white
    dots, anti-quarks as black and white dots. Dark shaded areas are
    due to the strong Coulomb fields at the particle positions and the
    contouring is left out here for better visibility of the flux
    structure.}
  \label{fig:energy-dens}
\end{figure*}

We expect that two strings of given length, that are asymptotically
far apart from each other ($D\gg R$), do only weakly interact. If they
approach gradually ($D\approx R$), parts of the flux tubes overlap and
both the scalar confinement field as well as the electric fields
are distorted from their asymptotic shapes. We show the energy
densities $\varepsilon(\vec{r})$ of these distorted fields in
fig.~\ref{fig:energy-dens}. We choose a string length $R=1\,$fm. In
the upper panel, the two strings are parallel to each other (type A)
and in the lower panel they are anti-parallel aligned (type
$B$/$B'$). From left to right the string distance $D$ decreases from
$D=1.5\,$fm to $D=0.5\,$fm. The symbols for the quarks are the white
dots, those for the anti-quarks the black and white dots. The contour
lines are equidistant in energy density in the range
$\varepsilon=(0.4\text{GeV/fm}^3\ldots 2.4\text{GeV/fm}^3)$ in steps
of $\Delta\varepsilon=0.4\text{GeV/fm}^3$. 

For large separations
($D=1.5\,\text{fm} > R$) there are two nearly unperturbed strings independent of
the orientation of the strings. In this case the string configurations
$A$ and $B'$ are indistinguishable. Only if the flux tubes get in contact
with each other ($D=1\,\text{fm}=R$), the two orientations behave
differently. For the parallel case, one finds only a slight attraction
between the flux tubes showing up in the distortion towards the center
of the two strings. Also the  energy density in the center of each
string is lowered a little bit. 

However, for the case of anti-parallel strings, 
the flux splits up in the two directions and is the same superposition
of types $B$ and $B'$ as already seen in fig.~\ref{fig:dfield-kappa}. 
We note, that this reorientation of the flux tubes is found just by
solving the equations of motion \eqref{eq:static_eoms}. No external
input such as the orientation of a Dirac-string as in the dual color
superconductor model \cite{Kodama:1997zc} is needed. The CDM therefore
incorporates the feature of string flip as used in the string-flip model 
\cite{Lenz:1985jk,Voss:1988hh,Horowitz:1991ux,Boyce:1993nt}. 
For the situation of anti-parallel strings, the string flip is not a
discontinuous process but a smooth transition. For large separations
($D=1.5\,\text{fm}$), there is already a small part of the
energy flux that stretches to the transverse direction. The same is
true for $D=0.5\,\text{fm}$ with interchanged roles of types $B$ and
$B'$. The 4-particle flux tube network is therefore a superposition of
types $B$ and $B'$ with continuously varying relative strength. 

If the two strings approach even further ($D=0.5\,\text{fm} < R$) the
two parallel flux tubes melt into an extended flux tube and form
a single bag (fig.~\ref{fig:energy-dens}, upper right panel). We
observe the contact of the two pairs of likewise 
charged particles which will lead to a repulsive Coulomb interaction.
For the anti-parallel situation however, the string-flip from type $B'$
to $B$ has nearly completed and the flux tubes point upside down
(fig.~\ref{fig:energy-dens}, lower right panel). The flux 
tube pattern now shows two flux tubes of length $D<R$. Because of this,
the definition of the strings is somewhat ambiguous. We start in
varying the distance $D$ between two $R=1\,\text{fm}$ long
$q\bar{q}$-strings and finally end up with two $q\bar{q}$-strings of
length 
$D$. Note that in this whole section we denote with the \emph{string
  length} the 
particle distance that we keep fixed, if not stated otherwise
explicitly.

\begin{figure}[htbp]
  \centering
  \includegraphics[width=\narrowfig,keepaspectratio,clip]%
  {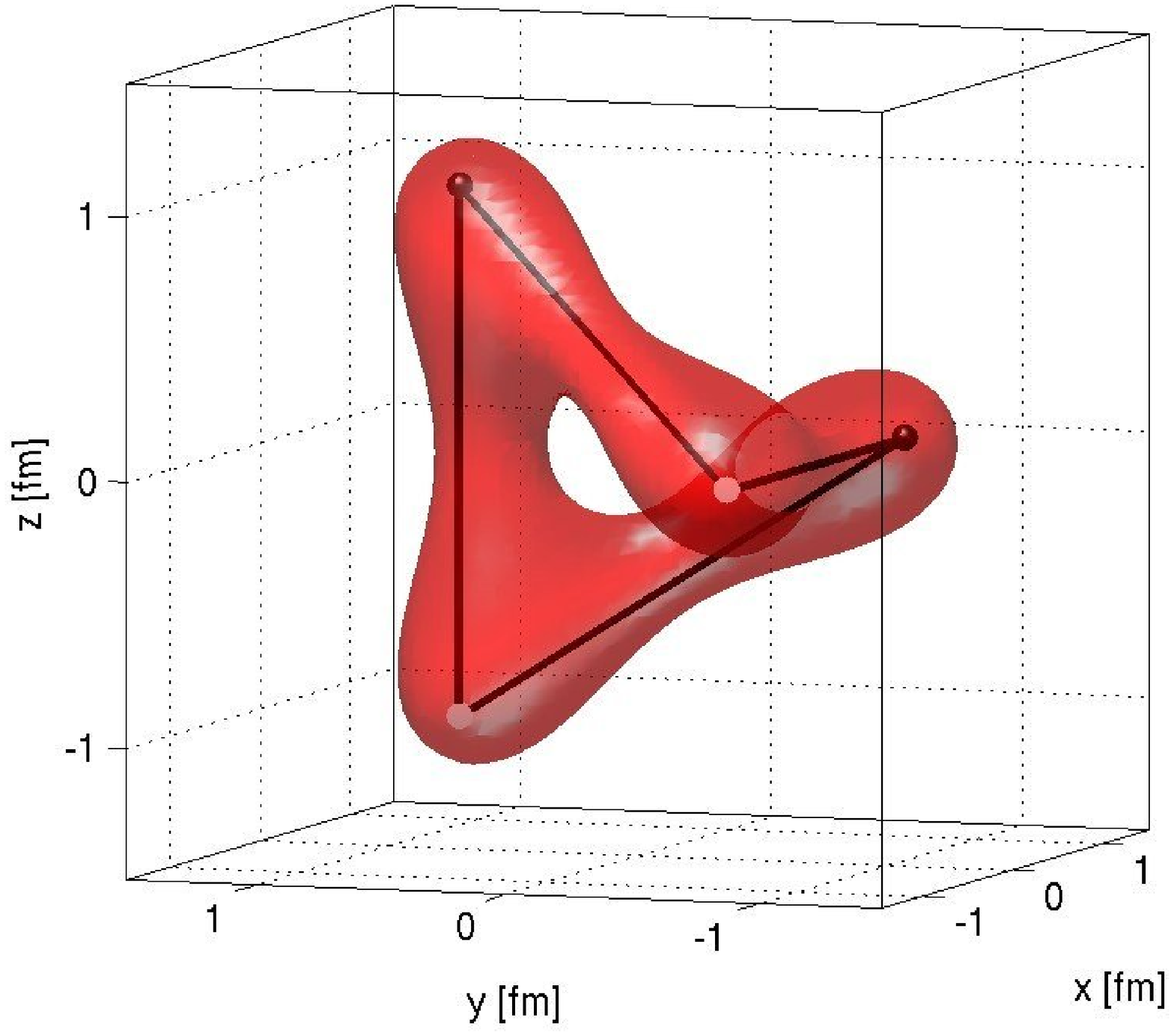}
  \includegraphics[width=\narrowfig,keepaspectratio,clip]%
  {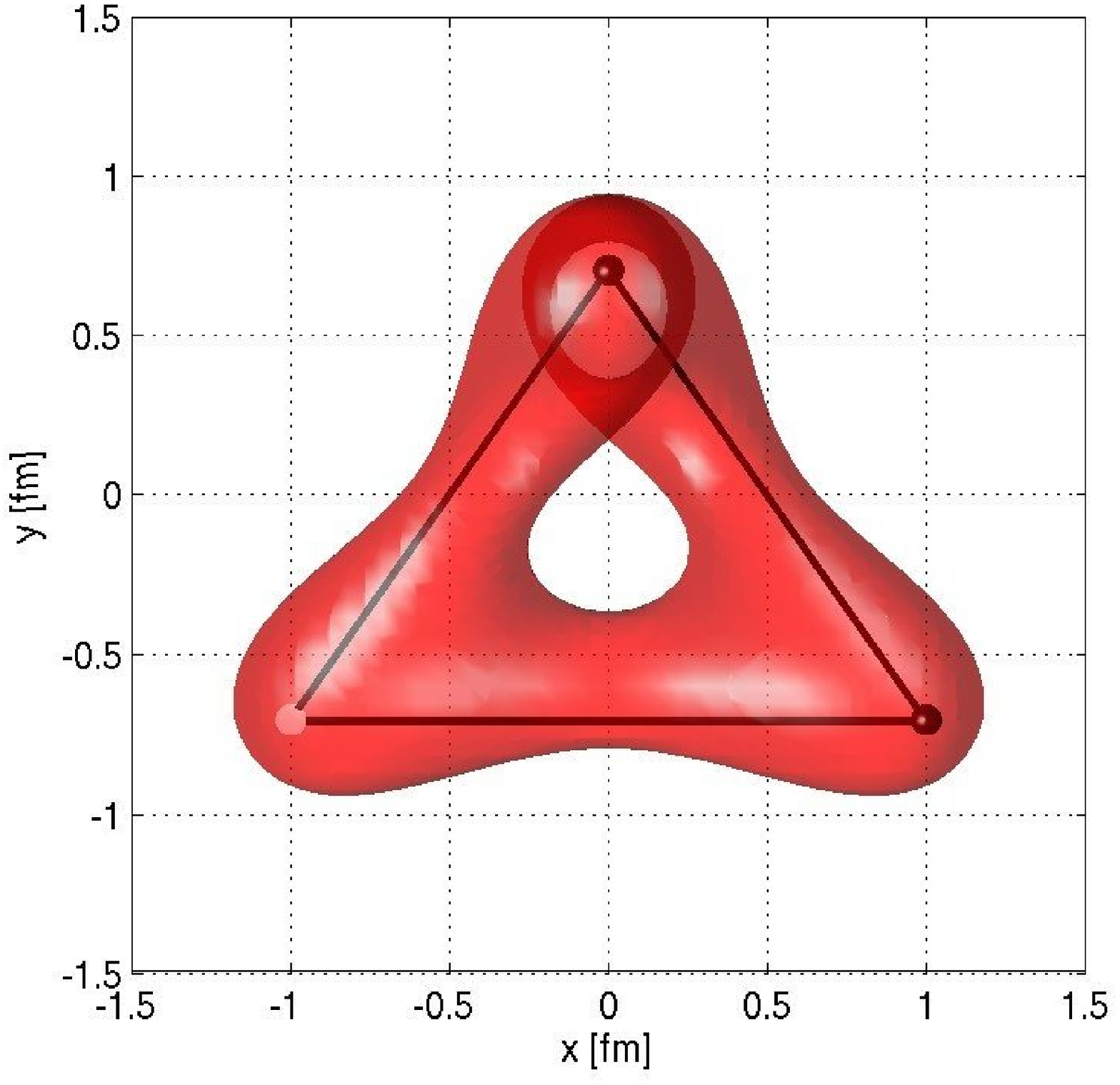}
  \caption{(color online). Four particles on  the corners of a
    tetrahedron. All pairwise 
    distances are equal to $R=2\,\text{fm}$. Shown is the
    equipotential surface $\kappa(\vec{r})=0.3$. Quarks and
    anti-quarks are marked as dark and light spheres,
    respectively. The lines are the direct links between each
    quark/anti-quark pair and are given to show the distortion of the
    interacting flux tubes. The upper and the lower plot give
    different views of the same configuration.} 
  \label{fig:tetrahedron}
\end{figure}

Of course, four particles do not have to be in a plane necessarily. As
our numerical realization does all calculations in three dimensions we
can easily describe flux tubes stemming from any arbitrary 3-dimensional
particle configuration. As one example we show the  flux tubes of
four particles placed on the corners of a tetrahedron with pairwise
distance $R=2\,\text{fm}$. We choose this large particle separation to
clearly show  the resulting flux tube network. In the 3-dimensional
illustration in fig.~\ref{fig:tetrahedron} the 
equipotential surface of the dielectric function $\kappa =
0.3$ is shown. In the center of a $q\bar{q}$-string $\kappa\approx
0.8$ for both parameter sets given in tab.~\ref{tab:params} (see
\cite{Martens:2004ad}). Therefore the value $\kappa=0.3$ is
characterizing the surface of the flux tube. The figure displays two
different views of the same configuration, the first a perspective
view of the system and the second a projection along the
z-axis. Quarks and anti-quarks are marked as black and white spheres, 
respectively. The four lines are the shortest links between each
quark/anti-quark pair. In this highly symmetrical configuration again
the flux of each quark is split up into two equal parts pointing
towards both anti-quarks, resulting in four equal flux tubes. The
centers of the flux tubes are bent 
slightly towards the center of the configuration, as already seen
before in the planar configuration. Moving one $q\bar{q}$-string out
of this symmetric configuration would cause two of the four flux tubes
to break in favor of the other two. The string flip is therefore seen
in the 3-dimensional configurations as well. 

\subsection{Interaction potentials}
\label{sec:potentials}

The energy of a $q\bar{q}$-system scales with the particle distance
$R$ according to the Cornell-potential given in
Eq.~(\ref{eq:cornell}). In the previous section we have shown the
energy distribution of two such interacting strings as well as the
difference of the 4-particle state to the incoherent superposition of
two equivalent $q\bar{q}$-strings. From the integral of the energy
density we get the total energy $E_4$ in Eq.~(\ref{eq:energy_decomp}) and
from the corresponding difference of energies in
Eq.~(\ref{eq:potential}) we extract the interaction potential
$V_4$. For the following calculation we keep the individual string
length $R$ and the relative orientation fixed, and vary only the
distance $D$ between the string centers.

\begin{figure}[htbp]
  \centering
  \includegraphics[width=\narrowfig,keepaspectratio,clip]%
  {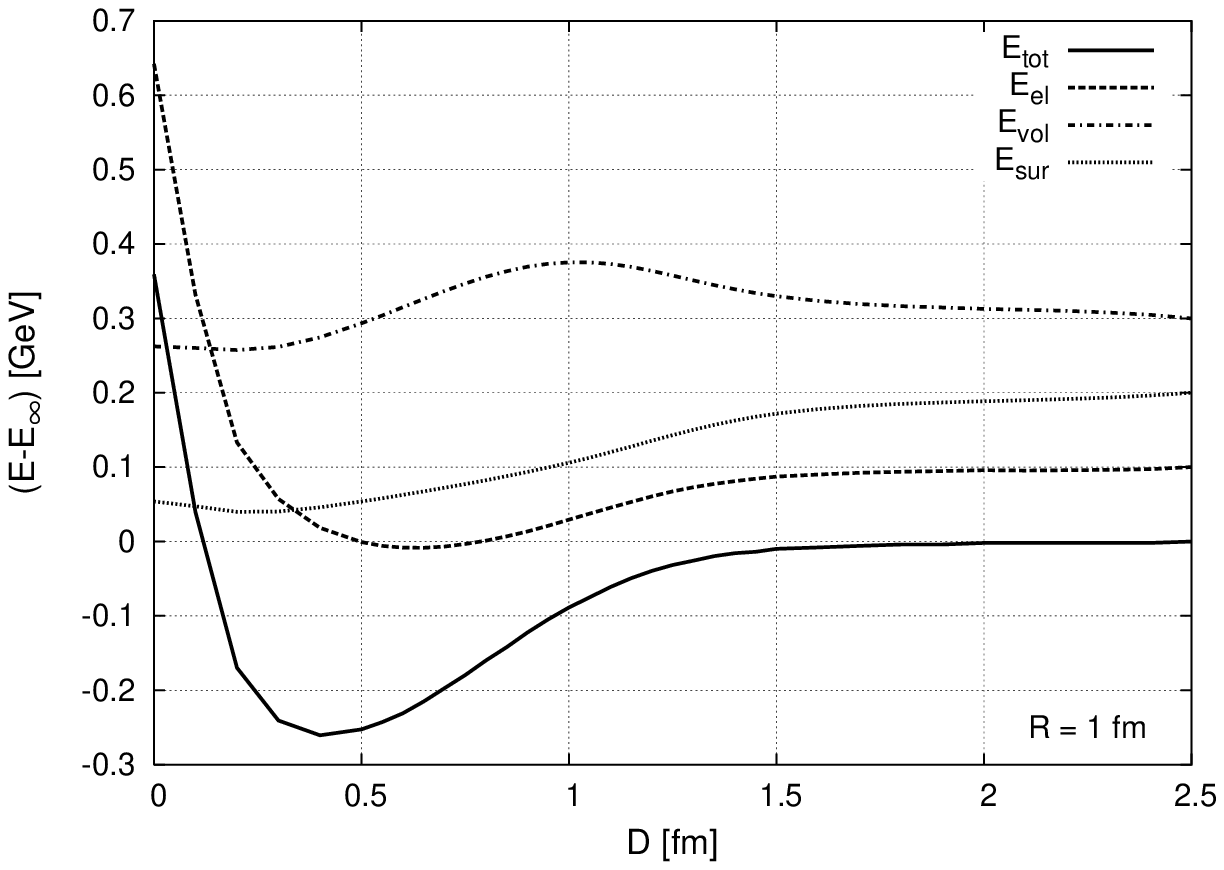}
  \includegraphics[width=\narrowfig,keepaspectratio,clip]%
  {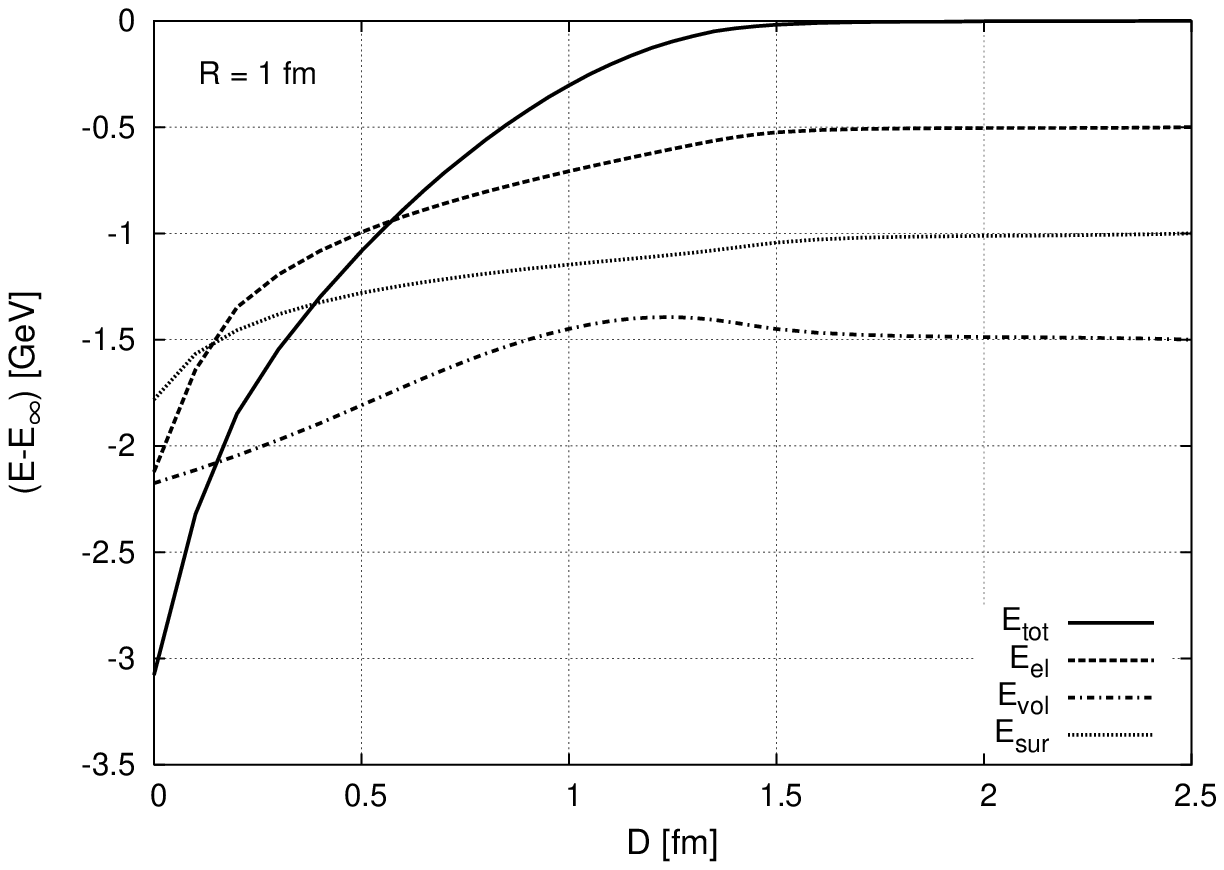}
  \caption{Energy $E_4$ of two parallel $q\bar{q}$-strings (top) and two
    anti-parallel $q\bar{q}$-strings (bottom) with length
    $R=1\,\text{fm}$. The total energy (solid line) is split into the
    different energy parts. All energies are reduced by their
    energies $E_\infty$ of two strings asymptotically far apart. The
    energy parts are separated by constant offsets of
    $0.1\,\text{GeV}$ (upper panel) and $0.5\,\text{GeV}$ (lower
    panel) for better visibility.}  
  \label{fig:strstr-energy-fracs}
\end{figure}

The total energy $E_4$ of two $R=1\,\text{fm}$ long flux tubes is
shown in 
fig.~\ref{fig:strstr-energy-fracs}. The orientation is parallel in the
upper panel and anti-parallel in the lower panel. The total energy
(solid line) is separated in the different energy parts according to
Eq.~(\ref{eq:energy_decomp}). Here we have subtracted from the total
energy the energy $E_\infty$ of two infinitely separated
$q\bar{q}$-strings. In the parallel case this is identical to the
potential $V_4$, as there is no string flip.
For clarity we have separated the curves for the three
energy fractions by equal offsets of $0.1\,\text{GeV}$ (upper panel) and
$0.5\,\text{GeV}$ (lower panel), respectively. 

The total energy saturates for distances $D>1.5\,\text{fm}$, i.e.~the
strings cease to interact. If they approach each other, we find for
the parallel strings a stable distance at $D\approx
0.4\,\text{fm}$. For smaller distances the two likewise charged
particles experience the Coulomb repulsion, which is seen only
in the electric part of the energy (dashed line). The volume part
of the energy (dashed-dotted line) exhibits a small maximum when the
two flux tubes get in contact. When they approach each other, the tails
of the strings overlap and the two-string configuration takes on a
larger effective volume than two isolated strings. As a consequence,
the volume energy rises while the electric energy drops down. 

For the anti-parallel strings the energy behaves similar for $D\apprge
1\,\text{fm}$. However, for smaller distances the string flip takes
place and the energy behaves nearly as two flux tubes that shrink
with $D$. 

\begin{figure}[htbp]
  \centering
  \includegraphics[width=\narrowfig,keepaspectratio,clip]%
  {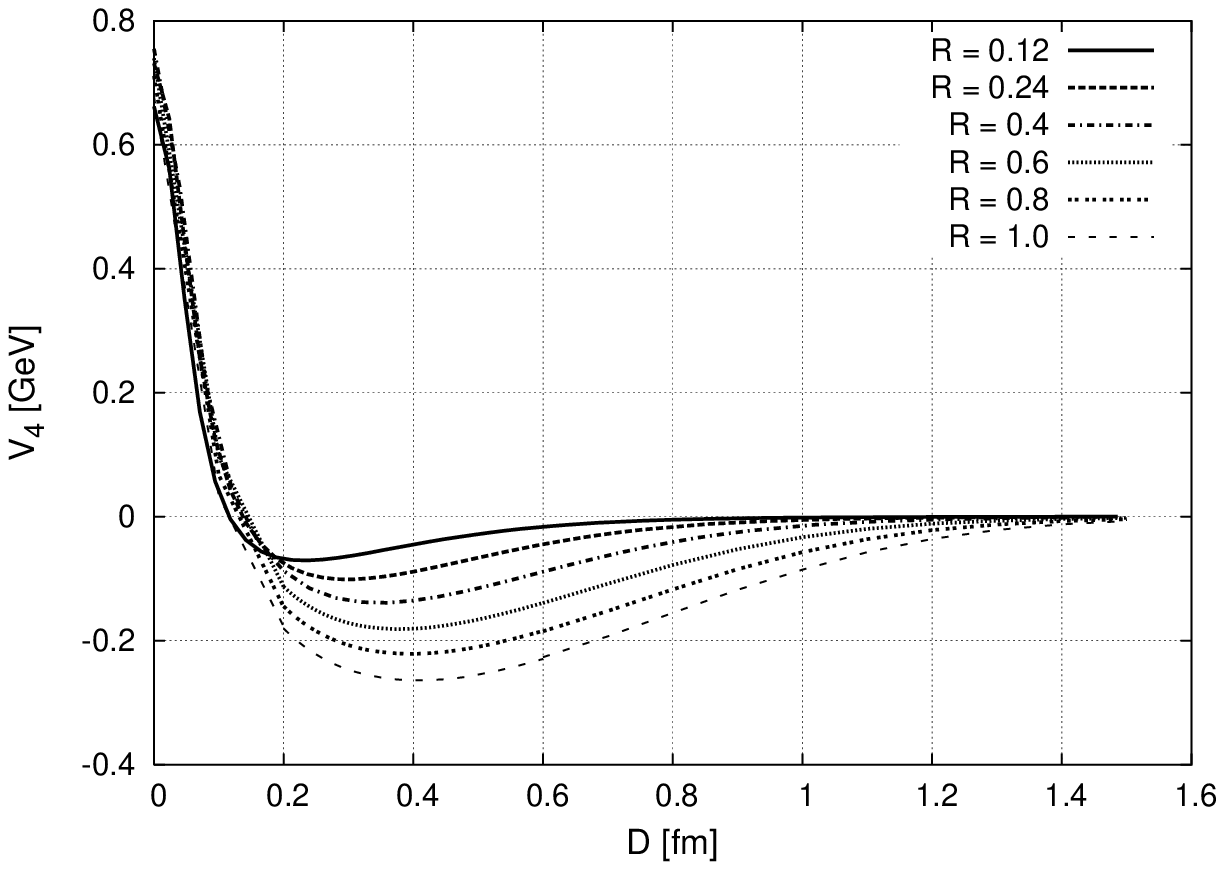}
  \includegraphics[width=\narrowfig,keepaspectratio,clip]%
  {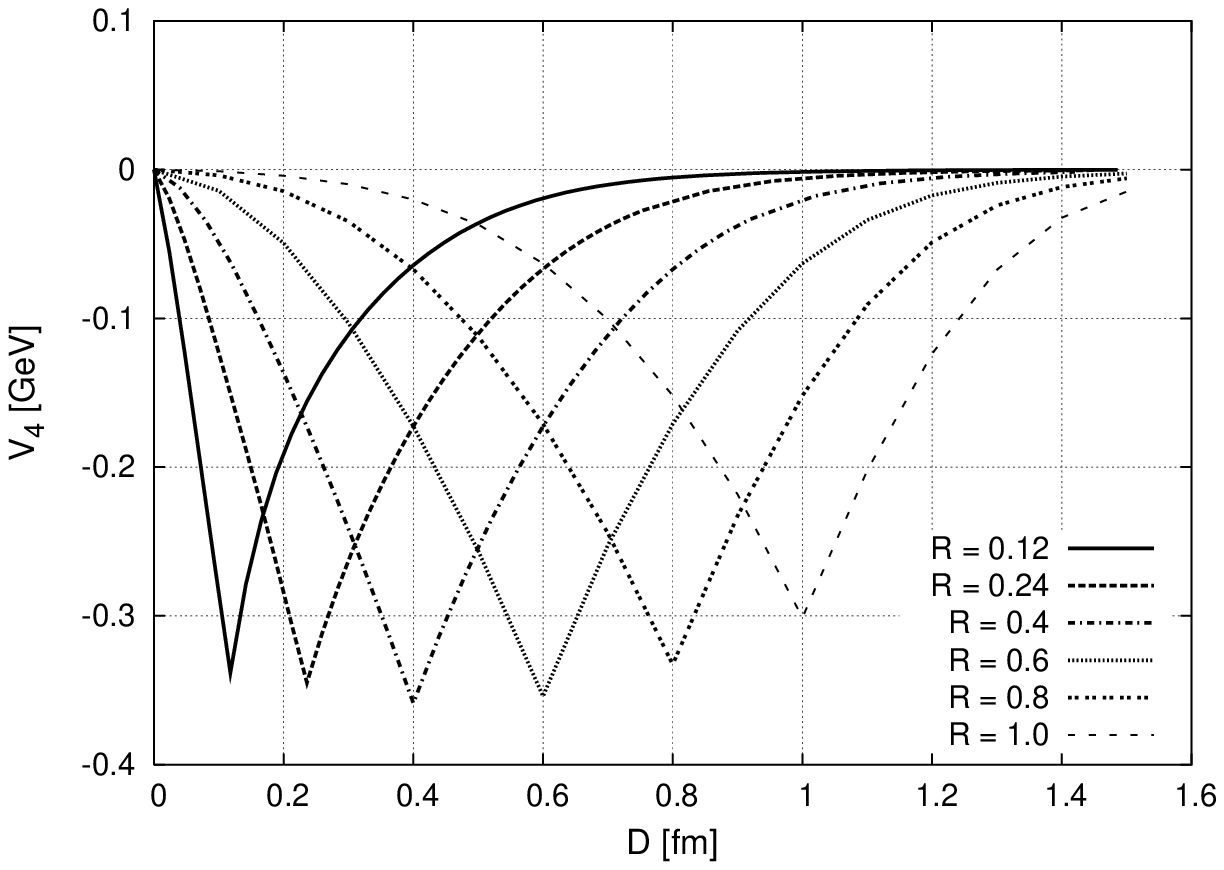}
  \caption{Interaction potential $V_4$ of two parallel $q\bar{q}$-strings
    (top) and two $q\bar{q}$-strings (bottom) with different lengths
    $R$. The kink in the anti-parallel configuration is due to the
    string flip at $D=R$.}  
  \label{fig:strstr-R}
\end{figure}

In fig.~\ref{fig:strstr-R} we present the interaction potential $V_4$
for different string lengths $R$. The orientation is parallel in the
upper panel and anti-parallel in the lower panel. The generic form
does not change with $R$, but both the position of the minimum and the
depth of the potential change with $R$. For the parallel strings the
minimum of the potential is moving with increasing $R$ to larger
separations $D$. For string lengths $R$ exceeding $1\,\text{fm}$ the
location of the minimum basically stays constant. It reaches a stable
point at $D\approx 0.4\,\text{fm}$ which is roughly the same size as
the radius of the single flux tubes (see
tab.~\ref{tab:params}). Qualitatively this potential resembles the
nucleon-nucleon interaction with a short/long range
repulsion/attraction and a dip of the order of 100 MeV.  

The potential of the anti-parallel strings exhibits a characteristic
kink at $D=R$, which is due to the string flip. At this point the
orientation of the strings flip from configuration $B$ to $B'$.
The results obtained in this work differ both qualitatively and
quantitatively from that in \cite{Loh:1997sk}. In the older work it
was assumed, that the electric fields of the to strings might be added
linearly, thus neglecting the 4-body interactions of the color
charges. In addition, the parameters chosen there led to flux tubes
with a radial width of $\rho_g \approx 1\,\text{fm}$, which is much
larger than our value $\rho_g \approx 0.4\,\text{fm}$ (cf.\
tab.~\ref{tab:params}). Consequently, the string interaction in the
work of Loh et.al.\ has a range  of about 1 fm. Also there was no
stable point in the potential as is seen in our work in
fig.~\ref{fig:strstr-R}.

\begin{figure}[htbp]
  \centering
  \includegraphics[width=\defaultfig,keepaspectratio,clip]%
  {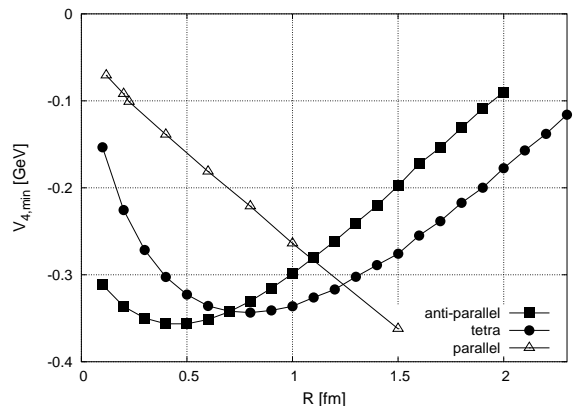}\hfill
  \caption{The minimum of the potential $V_4(D)$ for different 
    string lengths $R$. The magnitude of the potential
    $V_{4,\text{min}}$ rises linearly in the parallel case (open
    triangles) as expected for long homogeneous strings. In the
    anti-parallel case (solid squares) and the tetrahedron case
    (solid dots) the attraction gets weaker and shows an unexpected
    linear increasing behavior.}
  \label{fig:min-potential-R}
\end{figure}

In fig.~\ref{fig:min-potential-R} we have isolated the potential
minimum for different string lengths $R$. In the parallel case (open
triangles) the minimum scales linearly with the string length. This can
be understood, as for long parallel strings the profile does not
change along the flux tube axis. Thus the energy gain per string length
should be constant when the two strings get in contact.

For the anti-parallel configuration at $D=R$ but also for the
tetrahedron configuration, when all pairwise
particle distances are the same,  we can parameterize the
4-particle potential in the spirit of the 2-particle Cornell potential
$V_c(R)$ \eqref{eq:cornell}: 
\begin{subequations}
  \label{eq:4q-cornell}
  \begin{eqnarray}
    \label{subeq:4q-square}
    V_{\square} &=& 4 C_F E_0 
    - \left(\frac{4}{R} - \frac{2}{\sqrt{2}R}\right)\, C_F \alpha 
    + 4\tau_4 \, R - 2 V_c(R) \nonumber\\
    &=& - \left(2-\sqrt{2}\right)\,\frac{C_F\alpha}{R} + (4\tau_4 - 2\tau)\,R \\[1ex]
    \label{subeq:4q-tetra}
    V_{\text{tetra}} &=& 4 C_F E_0 
    -\left(\frac{4}{R} - \frac{2}{R}\right)\, C_F \alpha
    + 4\tau_4 \, R - 2 V_c(R) \nonumber\\
    &=& (4\tau_4 - 2\tau)\,R
  \end{eqnarray}
\end{subequations}
Here the first term in each bracket is due to the four attractive
$q\bar{q}$-pairs and the second to the two pairs $qq$ and
$\bar{q}\bar{q}$. The constant terms cancel each other exactly. 
The Coulomb interaction is only partially reduced in the square
configuration and completely canceled in the tetrahedron
geometry. 

In the above parameterization $\tau_4$ is an effective 4-particle
string tension, whereas $\tau$ denotes the standard $q\bar{q}$-string
tension from Eq.~\eqref{eq:cornell}. Naively one might estimate
$\tau_4$ in the following way. As we have 
seen in fig.~\ref{fig:energy-dens} (lower middle panel) and
in fig.~\ref{fig:tetrahedron}, the flux tubes of the anti-parallel
square configuration and of the  tetrahedron configuration do not
overlap for string lengths $R>1\,\text{fm}$. The electric flux from
each quark is therefore split into two equal flux tubes pointing to the
two anti-quarks. It was shown in \cite{Martens:2004ad}, that in the
CDM the string tension scales with $g_s\sqrt{C_F}$, i.e.~with the
charge of the particle. For the two symmetric configurations the
4-particle string tension therefore should be half of the
$q\bar{q}$-string tension $\tau_4 = \tau/2$. In this case the linear
confinement 
terms in Eq.~\eqref{subeq:4q-square} and Eq.~\eqref{subeq:4q-tetra}
vanish as well. 

In this simplified picture, the potential $V_\square$ of the planar
anti-parallel configuration should scale Coulomb-like and that for the
tetrahedron should vanish very rapidly, compared to the former
one. The result of the CDM calculations is shown in
fig.~\ref{fig:min-potential-R}.
For both the anti-parallel configuration (solid squares) and the
tetrahedron (solid triangles), the potential $V_{4,\text{min}}$ scales
linearly with the string length for $R>1\,\text{fm}$. Neither a
Coulomb-like nor a rapid cancellation is observed. Thus the 4-particle
potential is not a trivial combination of
$q\bar{q}$-potentials in the sense of the generalized Cornell
potentials as given in Eqs.~\eqref{eq:4q-cornell}.

The string-string potential
was also analyzed in SU(2) lattice theory for anti-parallel
configurations in \cite{Green:1993ag}. The qualitative behavior of the
potential is the same as in our model, although the absolute values
of the potential $V_4$ are consistently smaller than ours. However,
it should be noted that the parameters of our model used in this work
are not adjusted to the 4-quark 
problem, but are fixed to the $q\bar{q}$-properties only. In a very
recent SU(3) lattice calculation of the 4-particle system
\cite{Okiharu:2005ab} a multi-Y flux tube picture was proposed, such
that the total string length is minimized. This is similar to the
Y-like flux tube picture of the 3-quark system. In our model this
would be possible with another color content like
e.g.~$r\bar{r}g\bar{g}$, which is devoted to future work.

\begin{figure}[htbp]
  \centering
   \includegraphics[width=\defaultfig,keepaspectratio,clip]%
   {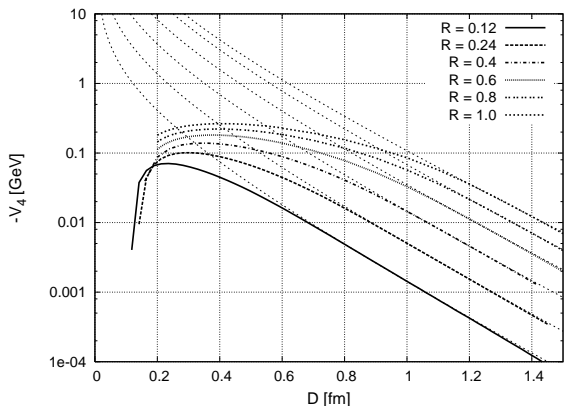}
  \caption{Logarithmic plot of the potential $V_4(D)$ for different
    string lengths $R$. The dotted lines are fits of a Yukawa type
    potential to $V_4(D)$. The model parameter correspond to PS1 and
    the strings are parallel to each other.}
  \label{fig:log-potential}
\end{figure}

\begin{figure}[htbp]
  \centering
  \includegraphics[width=\defaultfig,keepaspectratio,clip]%
  {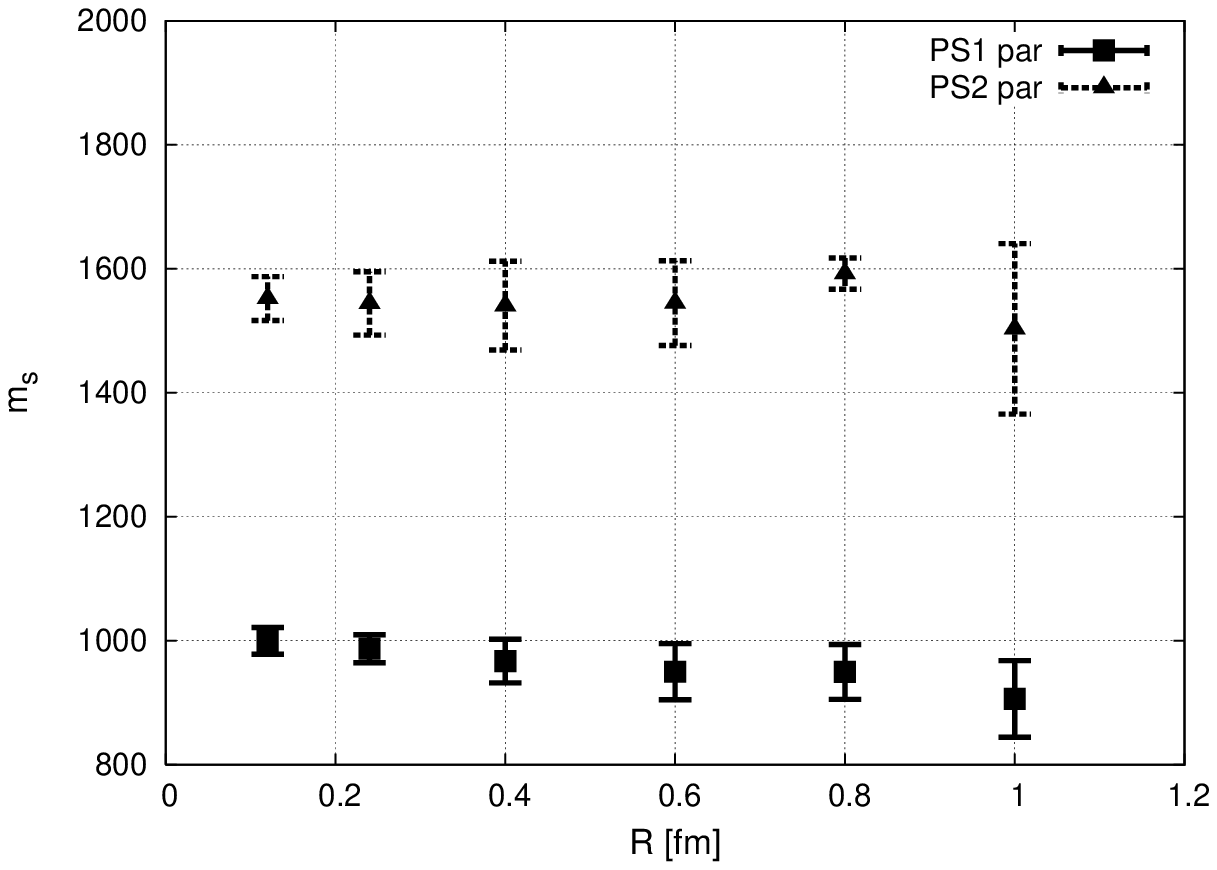}
  \includegraphics[width=\defaultfig,keepaspectratio,clip]%
  {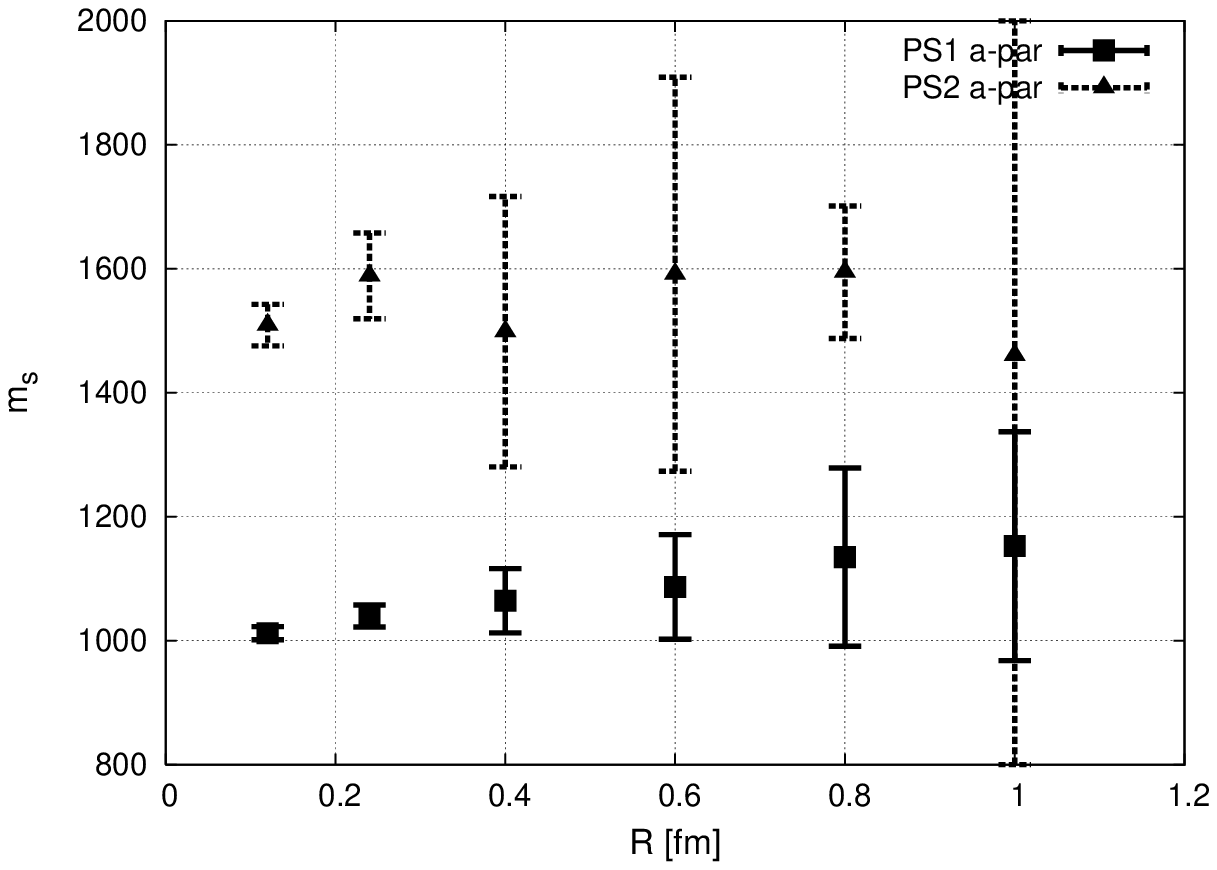}
  \includegraphics[width=\defaultfig,keepaspectratio,clip]%
  {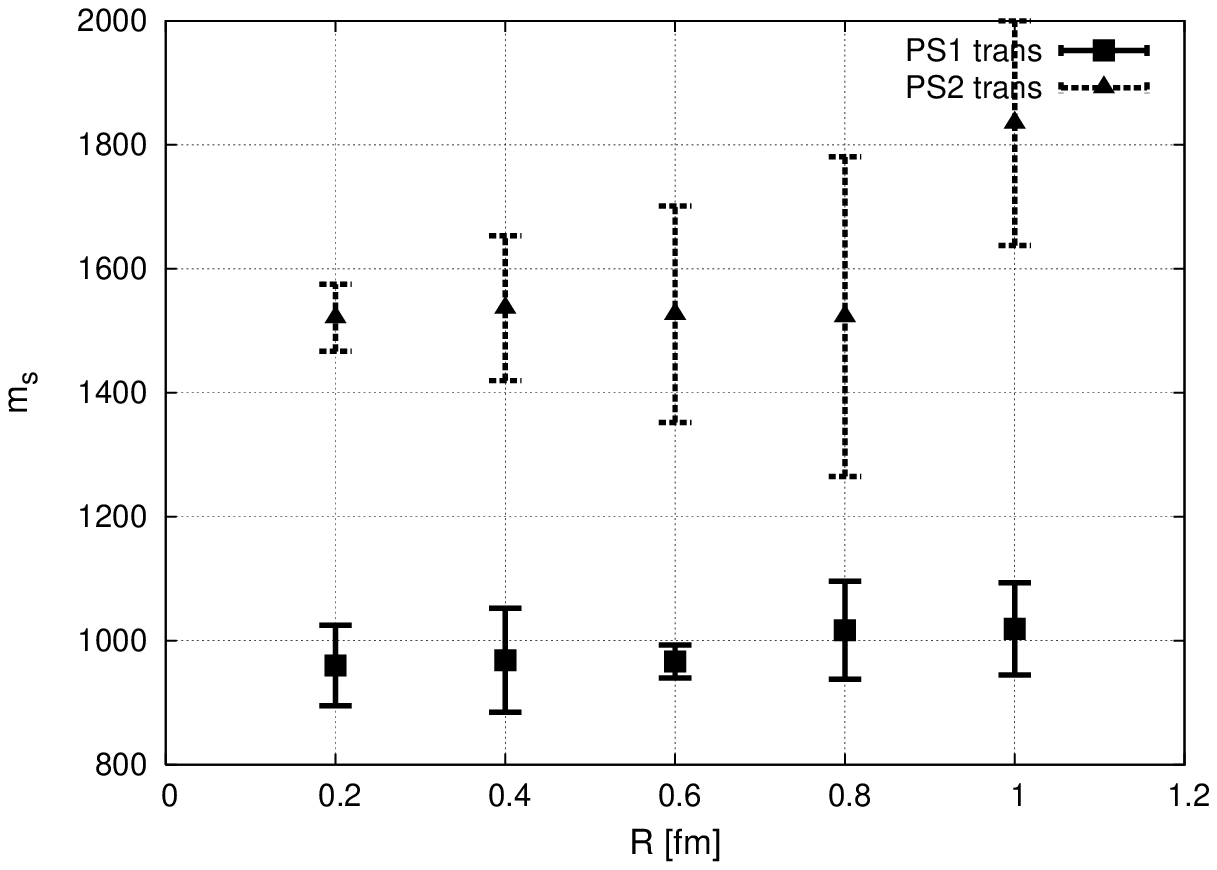}
  \caption{The screening mass of the interaction potential between two
    strings for PS1 (solid error bars) and PS2 (dashed error
    bars). From top to bottom the strings are oriented parallel, 
    anti-parallel and transverse (tetrahedron like) to each
    other. Within the uncertainties $m_s$ is equal to $m_g =
    1000\,\text{MeV}$ (PS1) and $m_g = 1500\,\text{MeV}$ (PS2).}
  \label{fig:screen_mass}
\end{figure}

Next we turn to the long range behavior of the potential
$V_4(D)$. From Eqs.~\eqref{eq:string-interaction} we expect a Yukawa
potential for sufficiently far separated strings. In
fig.~\ref{fig:log-potential} we show the negative of the potential
$V_4$ for different string lengths $R$ in a parallel orientation with
parameter set PS1. 
The dotted lines are the best
fits of a Yukawa type potential $V_\text{yuk}(D) = V_0 \exp\left(-m_s
  D\right)/(m_s D)$ as expected from Eq.~(\ref{eq:string-interaction})
to the CDM potentials, with $m_s$ and $V_0$ being 
fit parameters. For $D\apprge R$ the potential follows nicely the
Yukawa behavior. From Eqs.~\eqref{eq:string-interaction} we expect
also, that the screening mass $m_s$ is given by the curvature
$m_g$ of the scalar potential $U$ at $\sigma=\sigma_\text{vac}$. To
test this we show the dependence of $m_s$ on $m_g$ in
fig.~\ref{fig:screen_mass}. From top to bottom, the relative
orientation of the strings is parallel, anti-parallel and transverse
(tetrahedron like). We have extracted the screening mass from the fit
for different string lengths $R$ and for the two parameter sets PS1
and PS2 from tab.~\ref{tab:params}. The parameter $m_g$ takes on the
value $m_g = 1000\,\text{MeV}$ (PS1) and $m_g = 1500\,\text{MeV}$
(PS2), respectively. The error bars shown in the plot result from
variations of the fit interval of $D$ used in the fit. 
We note, that $V_4(D)$ is difficult to extract numerically, as it is
an exponentially small difference between to large energies (see
Eq.~\eqref{eq:potential}).  It should be
noted also, that it is numerically more difficult to calculate the
potential $V_4$ for large string lengths $R$ due to the limited
numerical box size but also for parameter set PS2 due to the strongly
pronounced maximum of the potential $U$ (see
fig.~\ref{fig:ups12}). Therefore the error bars are larger for longer
strings but also for parameter set PS2. However, within the errors a
good agreement of the screening mass to $m_s = m_g$ is seen. This
behavior is almost not dependent on the string length $R$ but also not
on the relative orientation.
In the dual Ginzburg-Landau model a Yukawa-type
potential was found with a screening mass $m_s = 1430\,\text{MeV}$
\cite{Kodama:1997zc}, which  is consistent with our results. Following
our above interpretation of $m_g$, one might compare these numbers to
the glueball mass which has been  calculated on the lattice between
$1500 \,\text{MeV} \le m_g \le 1700  \,\text{MeV}$
\cite{Morningstar:1997ff,Michael:1998tr}. Another possibility is
to compare $m_g$ with the mass $m_\text{off} \approx 1200 \,
\text{MeV}$ of the off-diagonal gluons given on the lattice as well
\cite{Amemiya:1998jz}. The detailed verification of the Yukawa
potential between two strings, as proposed in our description via the
CDM, should be a task for future lattice calculations.

\section{Multi-quark systems}
\label{sec:multi-quark}

In this section we present CDM results of overall colorless
multi-quark systems, with the particle number being larger than
four. Such states might in principle exist and a number of
possibilities like the pentaquark
\cite{Diakonov:1997mm,Nakano:2003qx}, the H-dibaryon
\cite{Jaffe:1976yi} and strangelets
\cite{Farhi:1984qu,Greiner:1988pc,Schaffner-Bielich:1996eh}, 
were eagerly discussed in the literature. Moreover we want to look for 
a possible transition if the quark number density becomes large as
e.g. in the interior of dense neutron stars or in the highly
compressed phase in relativistic ion collisions.
We study unordered ensembles of quarks and anti-quarks with
a varying number of particles. The particles are placed in a given
volume which is sufficiently smaller than the numerical box in order 
to reduce boundary effects. 

\begin{figure*}[htbp]
  \centering
   \includegraphics[width=0.32\widefig,keepaspectratio,clip]%
   {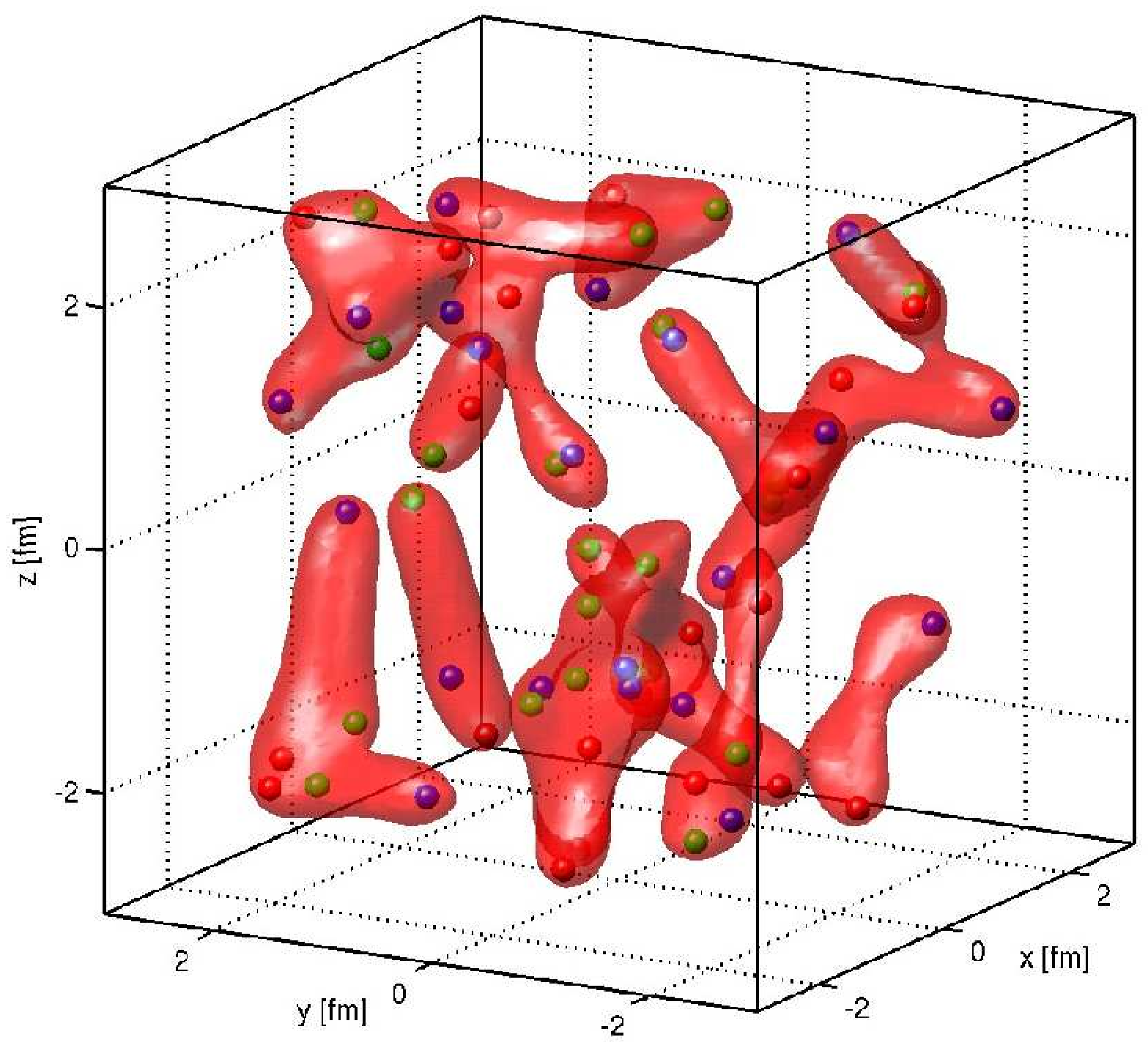} \hfill
   \includegraphics[width=0.32\widefig,keepaspectratio,clip]%
   {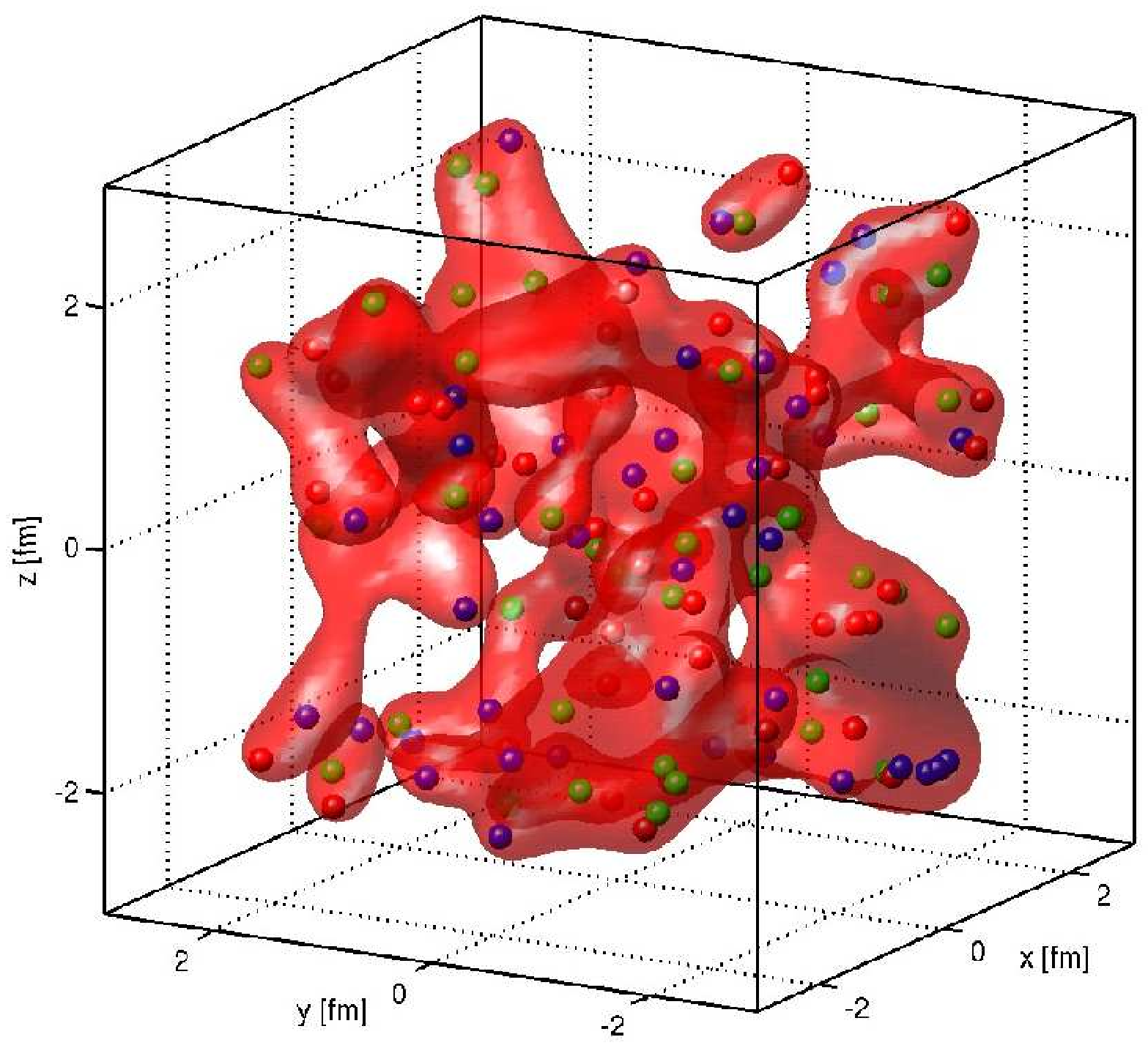} \hfill
   \includegraphics[width=0.32\widefig,keepaspectratio,clip]%
   {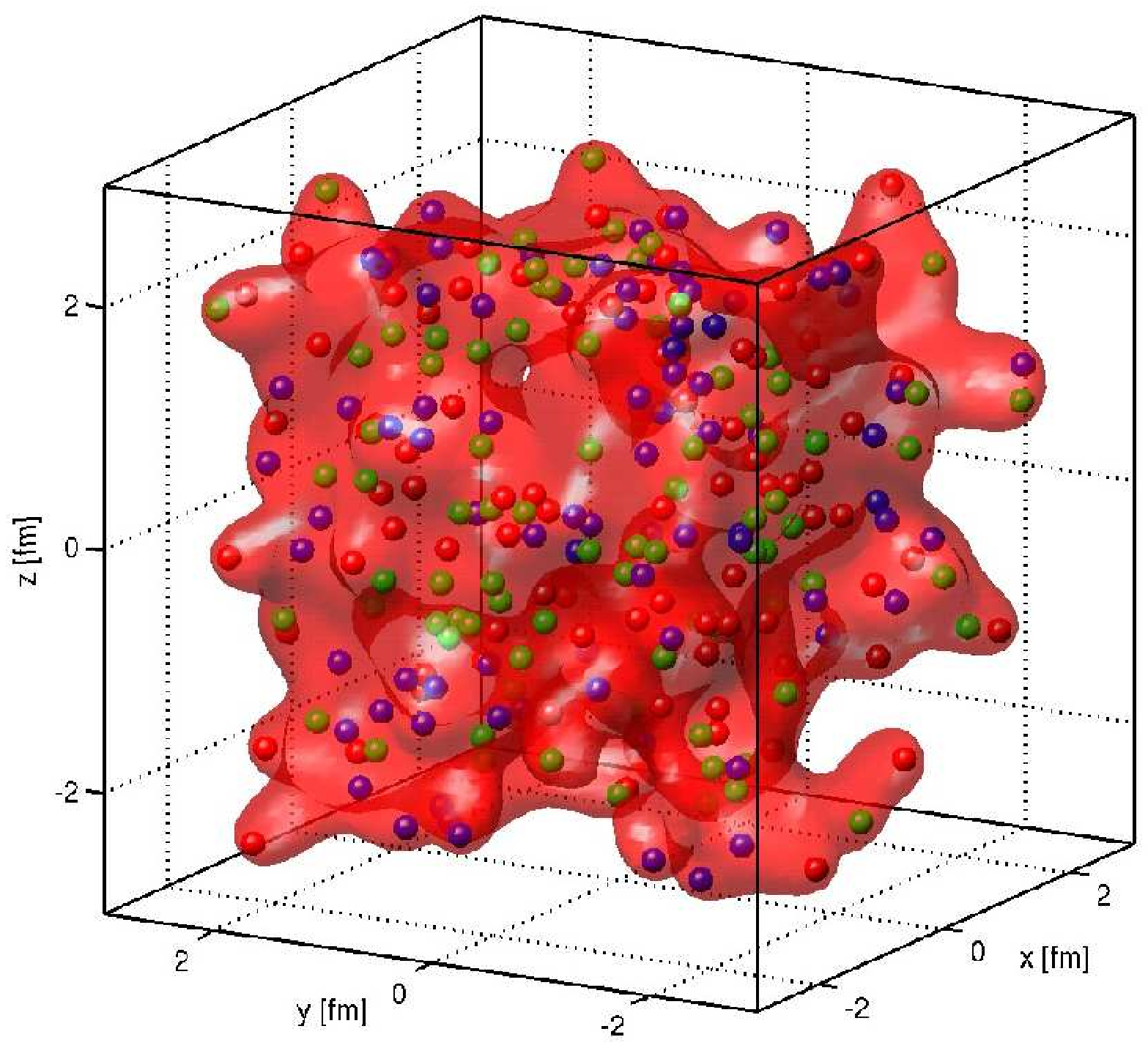}

   \includegraphics[width=0.32\widefig,keepaspectratio,clip]%
   {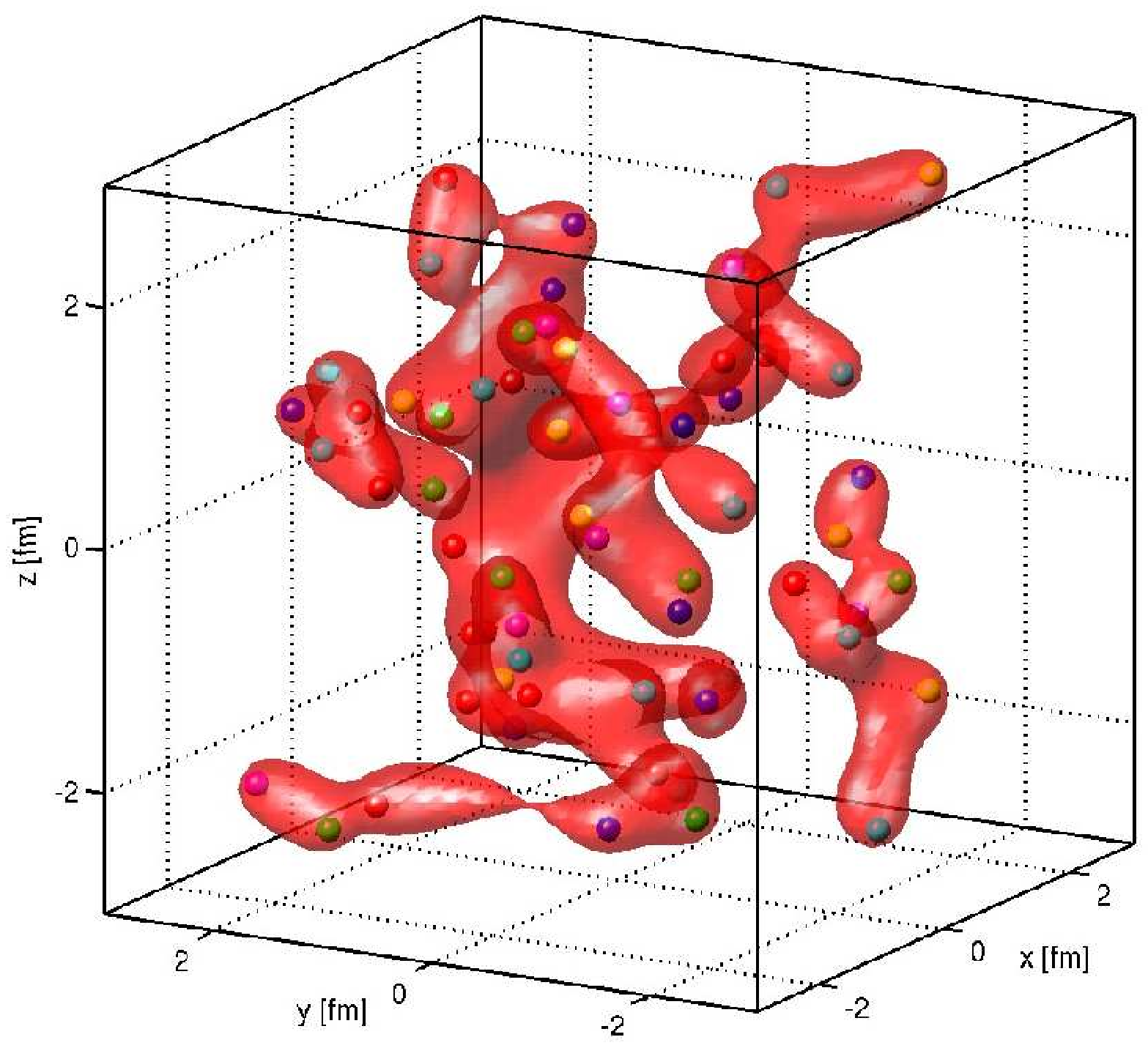} \hfill
   \includegraphics[width=0.32\widefig,keepaspectratio,clip]%
   {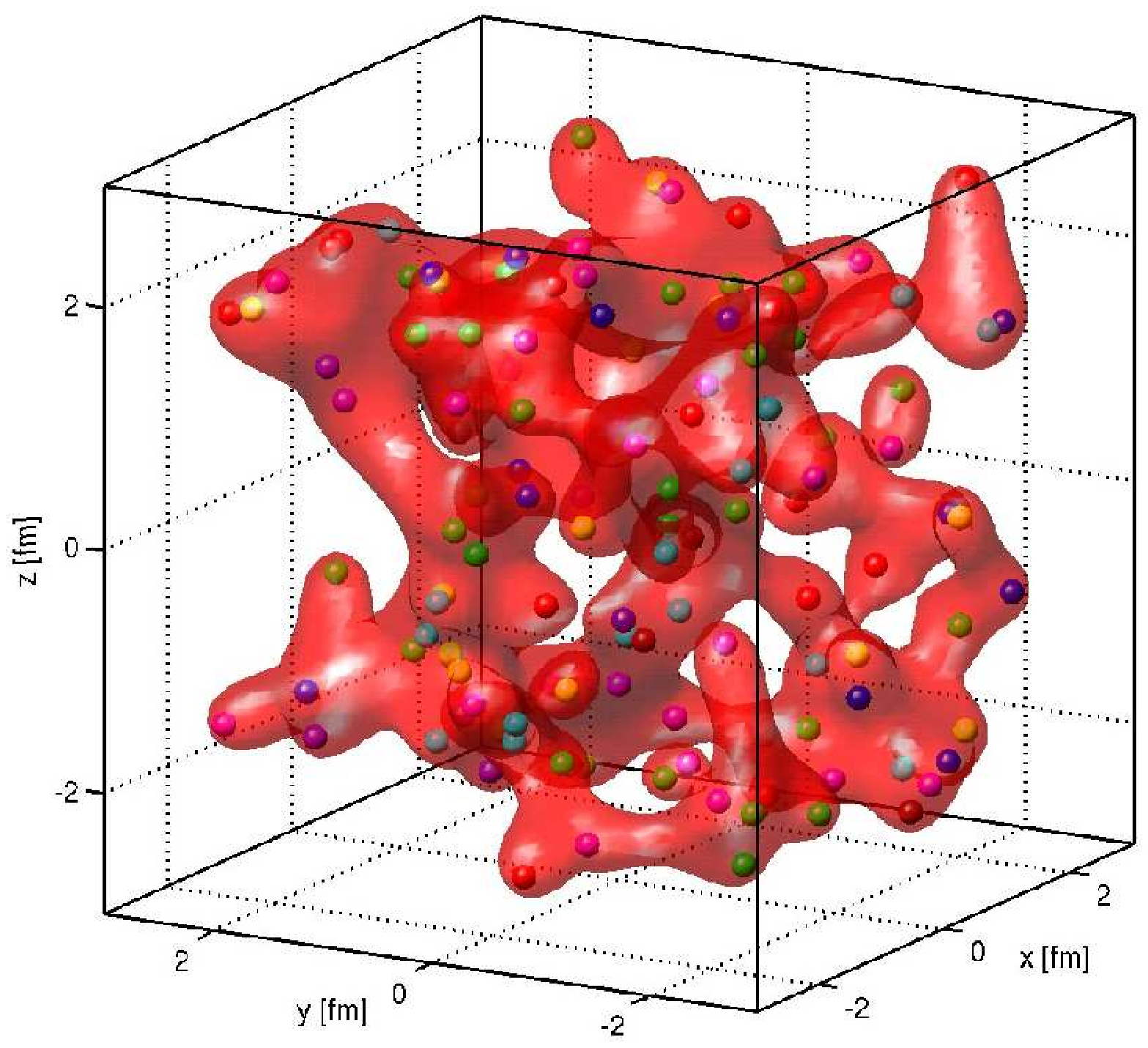} \hfill
   \includegraphics[width=0.32\widefig,keepaspectratio,clip]%
   {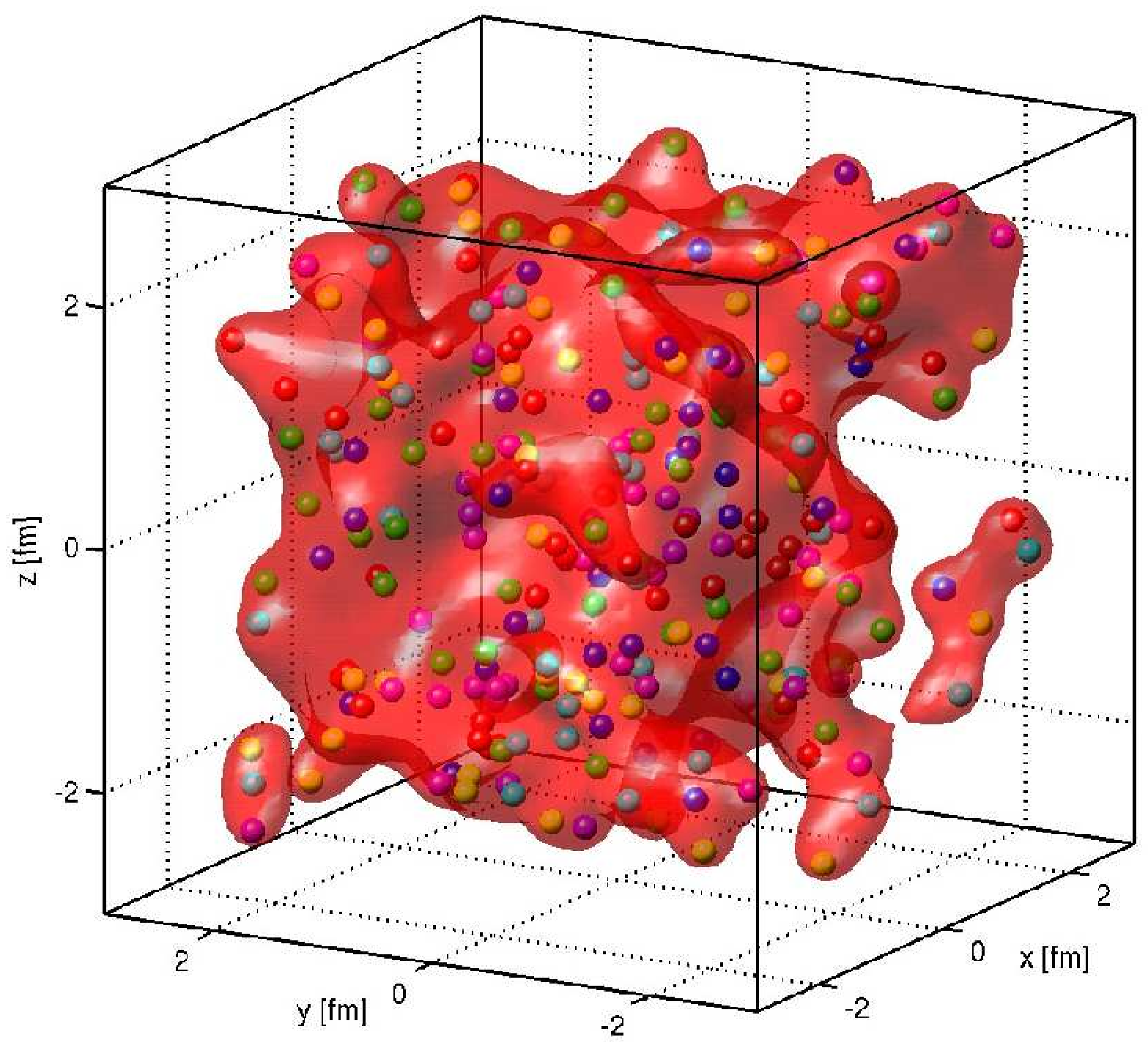}

  \caption{(color online). Flux tube structure of systems with baryon
    number $N_b>0$ (upper panel) and with vanishing baryon number
    (lower panel) for different particle densities. We show the
    equipotential surface of the dielectric function $\kappa_b = 0.4$
    characterizing the surface 
    of the flux tubes. Color codes for the particles are red/green/blue
    and cyan/magenta/yellow for the quarks and anti-quarks,
    respectively. The particle numbers for the upper row from left to
    right are $N=(63,126,255)$ and for the lower row $N=(64,128,256)$
    corresponding to particle densities $n = (0.5, 1.0,
    2.0)\,\text{fm}^{-3}$.} 
  \label{fig:multi-flux}
\end{figure*}
The color of the particles  are chosen
according to two different schemes. In the first 
we first pick a color $c\in \{r,g,b\}$. This color is assigned to a
quark $q$ and 
simultaneously the corresponding anti-color $\bar{c}$ is assigned to
an anti-quark $\bar{q}$. This $q\bar{q}$-pair is thrown randomly
into the volume. We repeat this procedure until a given number $N_m$ of
$q\bar{q}$-states is reached. In this way we get a system
with vanishing net baryon number, although baryonic and anti-baryonic
clusters might be formed. In the second scheme, we assign the
three colors $r,g,b$ to three quarks only and throw them into the
volume until a given number $N_b$ of baryonic $qqq$ states is
reached. In this scheme the baryon number is $N_b > 0$. 
Finally we vary the total number of quarks $N=2N_m$ and
$N=3N_b$, respectively, to study the behavior of the quark system for
different particle densities $n = N/V$. For each particle density $n$
we choose many different spatial configurations to calculate 
average quantities as the energy per particle and the bag volume per
particle. For this rather qualitative analysis we restrict ourselves
to the parameter set PS1. For percolation studies, such as the quark
density dependence of the mean bag size, or the formation of a single
super cluster, we would need more statistics. This analysis is devoted to
future work.

In fig.~\ref{fig:multi-flux} we show
the resulting flux tube structures for the baryonic (upper panel) and
the mesonic case (lower panel), respectively. In the latter one one can
find all possible color neutral subsystems,  namely $q\bar{q}$-states,
$qqq$-states and $\bar{q}\bar{q}\bar{q}$ as well as subsystems with
larger numbers of quarks and anti-quarks. For these figures the
particles were thrown into a cubic box with a size $V=(5\,\text{fm})^3$.
To visualize the flux tubes in
three-dimensional space, we show the position of the particles as
small spheres and the equipotential surface of the dielectric
function at a value $\kappa=\kappa_b=0.4$ characterizing the surface of the
flux tubes or bags. The number of quarks in the baryonic case (upper
panel) in
fig.~\ref{fig:multi-flux} from left to right is $N=63,126,255$
corresponding to particle densities $n = (0.5, 1.0,
2.0)\,\text{fm}^{-3}$ or baryon densities $n_b = n/3 = (1.0, 2.0, 4.0)
\times n_0$ with $n_0 = 0.17\,\text{fm}^{-3}$ being nuclear matter
density. The number of quarks and anti-quarks in the mesonic case
(lower panel) from left to right is $N=64,128,256$ leading
approximately to the same particle densities as in the baryonic case.

The formation of well defined bags is clearly seen. For small particle
numbers the system is dominated by long, nearly linearly shaped flux
tubes. In the baryonic ensemble the smallest clusters are
$qqq$-states, whereas in the mesonic ensemble the smallest clusters are
in principle $q\bar{q}$-states, $qqq$-states and
$\bar{q}\bar{q}\bar{q}$-states. For denser 
systems the  average number of particles per bag grows but still
distinct isolated bags are formed. Even for the $n=1.0\,\text{fm}^{-3}$
systems (the two centered figures) the flux tube
structure is maintained. The individual and still separated forms are
very complex objects and can be interpreted as multi-quark excitations
of hadronic particles much akin to the old bootstrap picture of higher
lying resonance states \cite{Hagedorn:1965st,Hagedorn:1967ua}. The
non-perturbative vacuum is replaced by a spaghetti like perturbative
vacuum and both vacua still balance each other. In the systems with
$n=2.0\,\text{fm}^{-3}$ (the two figures at the right) nearly all
particles are gathered in one single but highly deformed bag. The transition
from a small particle density with distinct bags and a small number of
particles per bag to a large particle density with only one
super-cluster and a large number of particles per bag is similar to a 
percolation transition which was proposed to occur for the quark
hadron transition \cite{Matsui:1986dk,Bauer:1988xxx,Nardi:1998qb}.

\begin{figure}[htbp]
  \centering
  \includegraphics[width=\narrowfig,keepaspectratio,clip]%
  {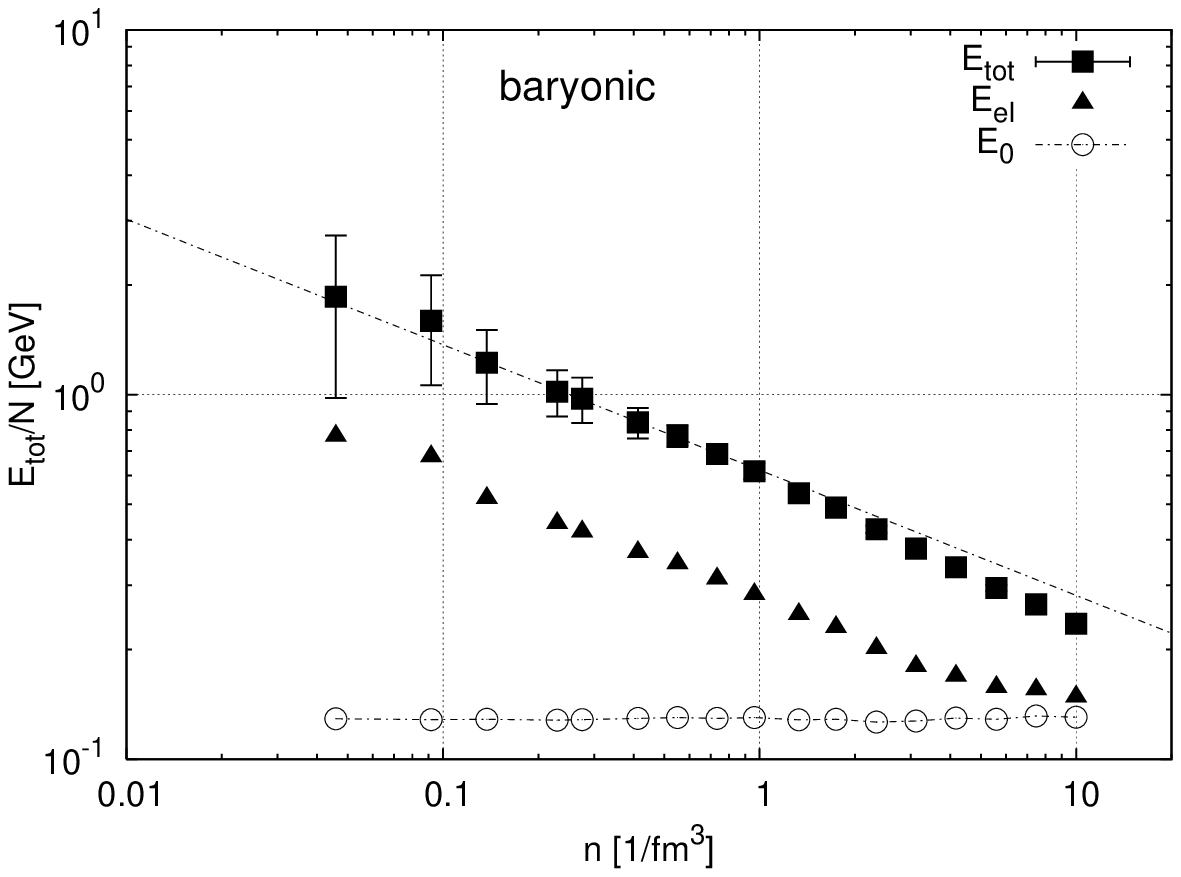}
  \includegraphics[width=\narrowfig,keepaspectratio,clip]%
  {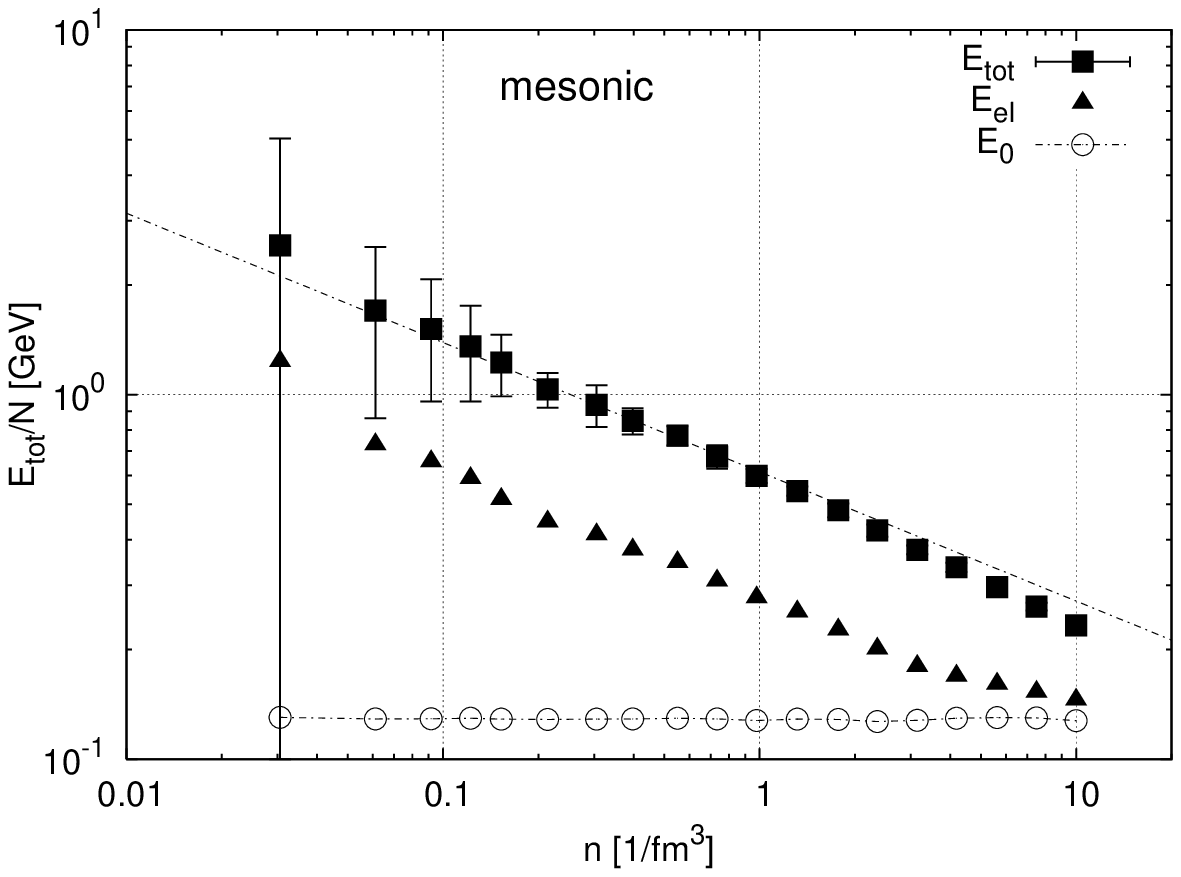}
  \caption{The energy per particle $E/N$ according to
    Eq.~\eqref{eq:energy_decomp} as a function of particle density
    $n$ for a baryonic (top) and a mesonic (bottom) system. Solid
    squares symbolize the average values at each density with the
    statistical standard deviations denoted by the error bars. Triangles
    are the average values of the 
    electric energy only and the open dots the average energy per
    particle for the Maxwell case ($\kappa=1$). The dashed line is a
    fit to the average energy per particle. }
  \label{fig:dc-energy-per-particle}
\end{figure}

In the following we analyze this interplay between the perturbative
and the non-perturbative vacuum more quantitatively. To reduce
numerical boundary effects, we reduce the volume, where the particles
are thrown in, to a spherical region 
with volume $V=\frac{4}{3}\pi r^3$ and radius $r = 2.5\,\text{fm}$. 
In fig.~\ref{fig:dc-energy-per-particle} we show
the energy per particle as a function of particle density in a double
logarithmic plot. The upper and the lower plot show a baryonic and a
mesonic system, respectively. The solid squares are the values for the
total energy per particle averaged over many configurations for each
density. The error bars denote the statistical standard deviation for
the ensembles of configurations. Clearly, the statistical
fluctuations of the energy is largest for the smallest densities. 
For comparison, we show the electric part of the energy 
(solid triangles) and the equivalent energy per particle $E_0/N$ for
the pure Maxwell case, i.e.~for $\kappa=1$ everywhere (open dots).
The fluctuations for the energy per particle are largest for small
densities $n$, as the particles get more and more homogeneously
distributed in space with growing density. Both the average total
energy and the electric energy follow roughly a power law, whereas the
free energy $E_0/N$ stays constant. We can estimate the scaling
of the total energy as follows. The energy of a cluster scales
linearly with the size $L$ of the cluster. For a given particle
configuration, the clusters form themselves by minimizing the total
energy. Each colored quark builds up a flux tube to the nearest
oppositely colored anti-quark or sub-cluster. Thus we can estimate the
cluster size by the average particle distance of the system,
i.e.~$L=n^{-1/3}$. The  total energy per particle is therefore
$E_\text{tot}/N \propto n^{-1/3}$. We have fitted the total energy per
particle to the  ansatz  $E_\text{tot}/N = c n^\beta$ in the range
$n \le 1\,\text{fm}^{-3}$. The result
for the fit parameter was $\beta=-0.35$ and $\beta=-0.36$
for the baryonic and the mesonic system, respectively, which is close
to the expected result $\beta=-1/3$. The fit to the low density region
overshoots slightly the high density results ($n > 3 \,
\text{fm}^{-3}$), indicating that the assumption of isolated long flux
tubes is not valid anymore. Instead, for very large densities, the
whole volume is filled with  particles 
homogeneously and the non-perturbative phase is pushed out of the
volume $V$. The dielectric function is therefore $\kappa \apprle 1$
everywhere inside $V$ and the electric energy should be the same as for the
equivalent free case. Indeed, the electric part of the energy
(triangles) approaches  the free energy (open dots) for densities $n >
5\,\text{fm}^{-3}$. 

\begin{figure}
  \centering
  \includegraphics[width=\defaultfig,keepaspectratio,clip]%
  {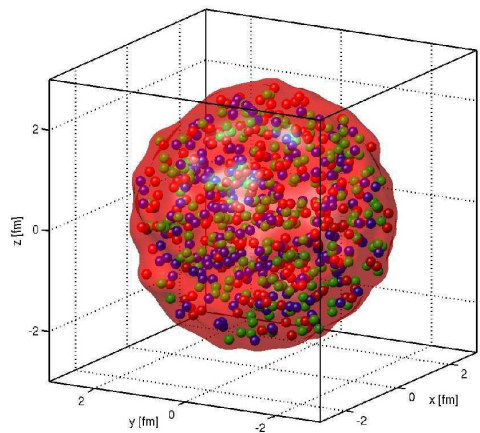}
  \caption{(color online). 654 quarks randomly distributed in a sphere
    with radius $r=2.5\,\text{fm}$ and volume $V$. The
    non-perturbative phase ($\sigma=\sigma_\text{vac}$) is pushed out
    of the sphere completely and $V_\text{pert}\approx V$. The bag
    surface is defined by $\kappa = \kappa_b = 0.4$.}
  \label{fig:quark-blob}
\end{figure}
Another interesting quantity is the specific bag volume per particle
$v_\text{pert} = V_\text{pert}/N$. For small $n$ one finds dominantly
bags with two 
or three particles. The size  of the bags per particle decrease with
the average particle distance,  i.e.~with increasing $n$, until with
still further increasing density the bags start to overlap
and melt together. The decrease of the bag volume per particle slows
down or might even turn over to an increase with increasing $n$ in the
same way as we have seen in the bump of the volume energy for the
melting $q\bar{q}$-system in fig.~\ref{fig:strstr-energy-fracs}.
At a critical density $n_c$, when the melting of the bags is completed
and the non-perturbative phase is pushed out of the volume $V$ completely,
the total perturbative volume can not increase anymore and
$V_\text{pert}=\text{const}=V$ as shown in fig.~\ref{fig:quark-blob}.
With further increasing density
$v_\text{pert}$ decreases as $n^{-1}$. This critical density $n_c$
marks the transition to a system of deconfined quarks and anti-quarks.

\begin{figure}[htbp]
  \centering
  \includegraphics[width=\narrowfig,keepaspectratio,clip]%
  {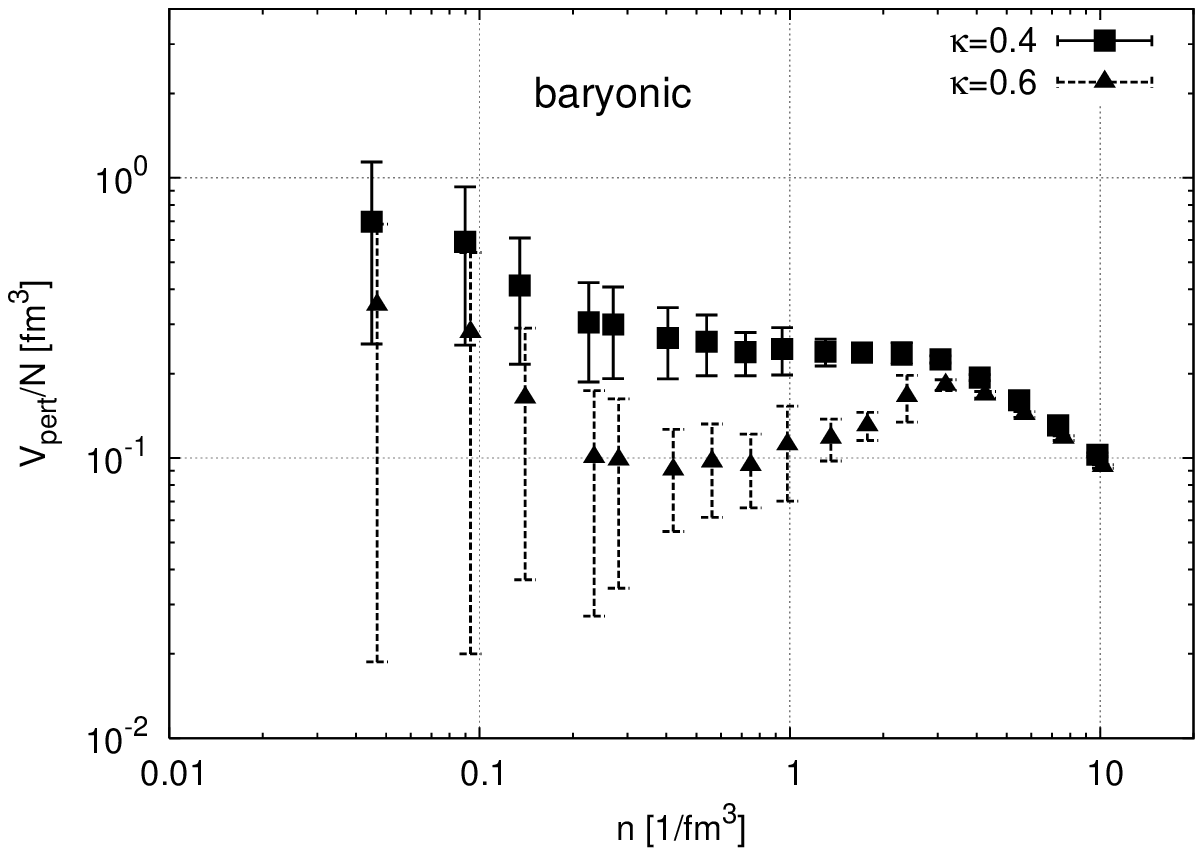}
  \includegraphics[width=\narrowfig,keepaspectratio,clip]%
  {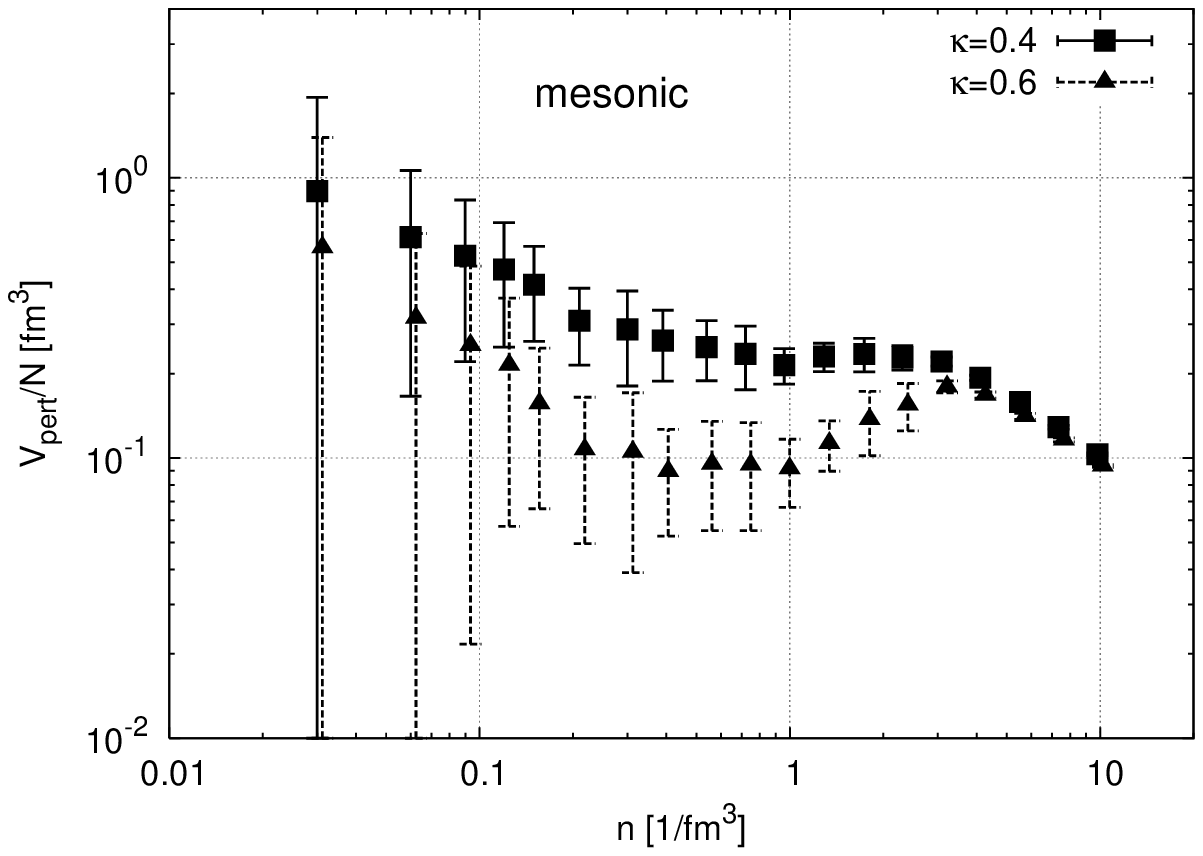}
  \caption{The specific bag volume per particle for a baryonic (top) and
    mesonic system (bottom). Squares denote the average specific bag
    volume for $\kappa_b=0.4$ and triangles that for
    $\kappa_b=0.6$. Error bars indicate the standard deviation for the
    set of different configurations at each density. At $n_c \approx
    3\,\text{fm}^{-3}$ the melting of the single bags is
    completed and the quarks are in the deconfined phase.}
  \label{fig:dc-perturbative-volume}
\end{figure}
We measure the total bag volume as that part in space, where $\kappa
\ge \kappa_b$. To show the dependence of the bag volume on $\kappa_b$
we choose the two values $\kappa_b=0.4$ and $\kappa_b=0.6$, where
$\kappa_b = 0.4$ ($\kappa_b=0.6$) is a surface 
somewhat more in the exterior (interior) of the bag. Therefore the bag
volume measured in this way is larger for $\kappa_b=0.4$ than for
$\kappa_b=0.6$. The  result is shown in
fig.~\ref{fig:dc-perturbative-volume} for the  baryonic (top) and the
mesonic case (bottom) for $\kappa_b=0.4$ (squares) and
$\kappa_b=0.6$ (triangles). Again the error bars indicate the standard
deviation of the specific volume for the different measured
configurations at each density. The bag volume per particle
shows exactly the anticipated result. It decreases with increasing $n$
for small densities. For $0.3\,\text{fm}^{-3} \le n \le
3\,\text{fm}^{-3}$ the specific bag volume $v_\text{pert}$ develops
a plateau for $\kappa_b=0.4$ and rises again for $\kappa_b=0.6$. We
find the critical density $n_c$, i.e.\ the point where
$v_\text{pert}$ decreases again, at around $n_c\approx
3.0\,\text{fm}^{-3}$. This corresponds to a baryon density $n_b =
n_c/3 \approx 6 n_0$, i.e.\ six times nuclear matter density or a
meson density of approximately $n_m = n_c/2 = 1.5\,\text{fm}^{-3}$.
We note,
that even at the highest density $n=10\,\text{fm}^{-3}$ the mean
particle distance is much larger than the intrinsic particle size $r_0$ of the
particles, so that the color charges do not cancel out each other by
overlapping. 

\begin{figure}[htbp]
  \centering
  \includegraphics[width=\defaultfig,keepaspectratio,clip]%
  {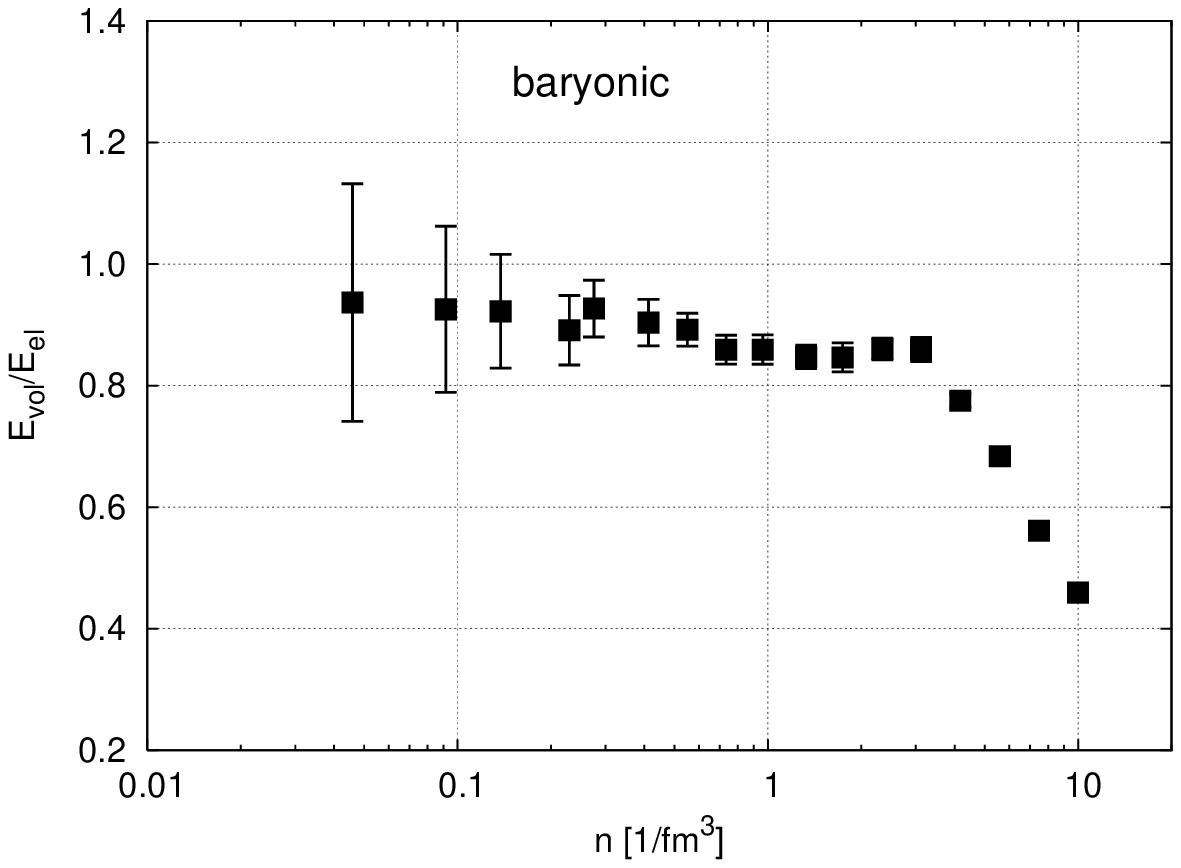}
  \includegraphics[width=\defaultfig,keepaspectratio,clip]%
  {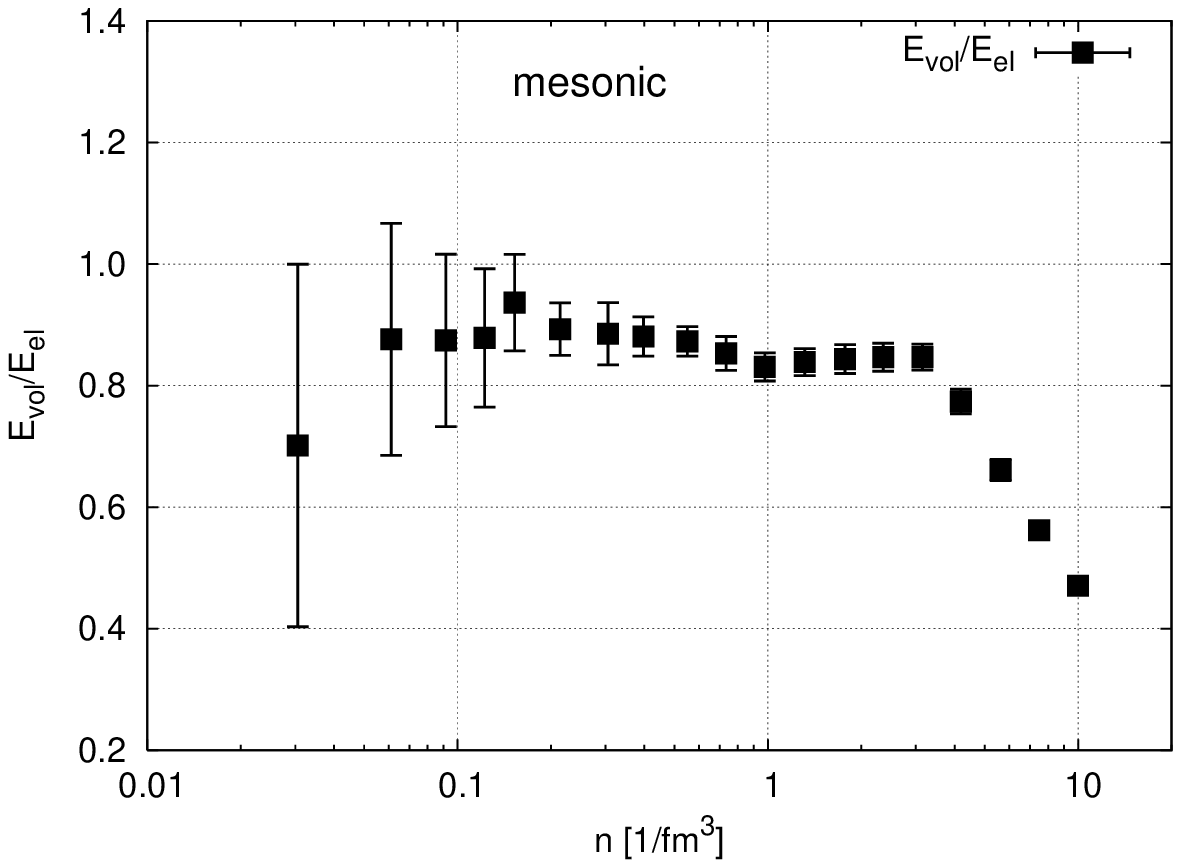}
  \caption{The balance between the electric and the volume energy for
    a baryonic (top) and a mesonic system (bottom). The open squares
    and the solid dots denote the values of every configuration and
    their average value, respectively. The two energies
    are the  same within 10\% for the numerically reasonable range.}
  \label{fig:dc-energy-ratio}
\end{figure}
It was one of the results of the work in \cite{Martens:2004ad}, that
the volume energy $E_\text{vol}$ and the electric energy $E_\text{el}$
of a $q\bar{q}$-string nearly balance each other. This holds for larger
particle densities as well until the system reaches the deconfinement
transition. In fig.~\ref{fig:dc-energy-ratio} we show
the ratio $E_\text{vol}/E_\text{el}$ as a function of the particle
density for a baryonic (top) and a mesonic system (bottom). It is constant
within 10\% (neglecting the first point with large uncertainties in
the mesonic system) for 
densities $n \apprle 3.0\,\text{fm}^{-3}$ but falls down for higher
densities once the system is in the deconfined phase. The bag volume
and therefore the volume energy $E_\text{vol}$ are maximal but the
electric selfenergy associated with the particles increases with the
particle density. 

\begin{figure}
  \centering
  \includegraphics[width=\narrowfig,keepaspectratio,clip]%
  {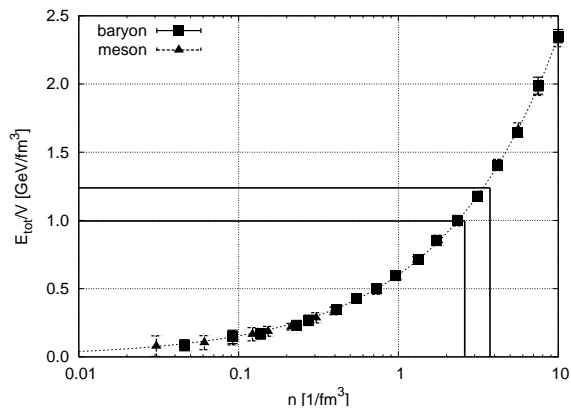}
  \caption{Energy density as a function of particle density
    $n$. Baryonic (squares) and mesonic systems (triangles) are
    indistinguishable from each other. The energy density is well
    described by a power law (dashed line) and is a smooth function
    even across the critical density at $n_c =
    3.0\,\text{fm}^{-3}$. The vertical and horizontal lines mark the
    uncertainties in the critical particle density and critical energy
    density, respectively.}
  \label{fig:energy-density}
\end{figure}
Finally we show the total energy density
$\varepsilon=E_\text{tot}/V$ as a 
function of the particle density in fig.~\ref{fig:energy-density} for
both baryonic (squares) and mesonic (triangles) systems. It follows
nicely a polynomial behavior (dashed line) over the whole density
range. Remarkably, it is indistinguishable for (static) mesons and
baryons. At the critical density $n_c = 3.0 \, \text{fm}^{-3}$ it has a
value $\varepsilon_c = 1.2 \, \text{GeV}/\text{fm}^3$. Due to
the uncertainty in the exact value of the critical density $n_c$ we
give a band for the critical energy density between
$1.0\,\text{GeV}/\text{fm}^3$ and $1.3\,\text{GeV}/\text{fm}^3$. This is in good
agreement with the critical energy density found on the lattice
\cite{Karsch:2004ti}, where it was found to be between
$0.5\,\text{GeV}/\text{fm}^3$ and $1.0\,\text{GeV}/\text{fm}^3$.

We note, that there are basically no differences in
figs.~\ref{fig:dc-energy-per-particle},
\ref{fig:dc-perturbative-volume} and \ref{fig:dc-energy-ratio} between
baryonic and mesonic systems. Apparently all effects are driven by the
particle density, no matter which kind of particles (quarks and
anti-quarks or only quarks) are studied.

\section{Summary and discussion}
\label{sec:summary}

The description of the interactions of quarks and gluons, both at high
temperatures/densities and in vacuum, from first principles is still
an open task to solve. The chromodielectric model offers a tool to
describe in a transparent way these interactions while including the
confinement phenomenon dynamically within the same framework.

Within the chromodielectric model particles group
themselves into color neutral bags with the means of color electric
flux tubes. In this work we studied the interactions of these flux
tubes with each other. We have seen both on the level of the electric
fields and on the level of the energy density, that the flux tubes
attract each other. Depending on the relative orientation of the
strings, the flux tubes either melt gradually together or change the
direction of the electric flux via a characteristic string flip. This
should also be seen in future lattice SU(3) calculations for many heavy
quarks. 

The attraction is seen also in the interaction potential between the
strings. The depth as well as the range of the interaction changes
with the length of the strings. The interaction depth in the order of
a few hundred MeV is relatively strong compared to a rather old analysis
made in lattice SU(2) calculations but also compared to a similar
analysis within the dual color superconductor model. In the case of
the string flip situation  the potential can not be described by
a simple incoherent superposition of flux tubes, but shows a real
multi-particle effect. The long range behavior of the potential can be
well described by a Yukawa type potential. The exponential decay of
the interaction is expressed by a screening mass between the flux
tubes which is rather insensitive to the relative orientations of the
strings and to their length. The screening mass  scales
with the mass parameter $m_g$ introduced in the scalar
self-interaction of the confinement field. It is a future task to
check the model continuously against upcoming results performed on the
lattice. On the other hand the Yukawa interaction should be tested in
detailed future four-quark lattice calculations. Only recently we have
become aware of a lattice study about the interaction of four- and
five-quark  systems \cite{Okiharu:2005ab,Okiharu:2005eg}.  

In the studies of the multi-particle systems we have explored the
structure of the corresponding flux tubes. The non-perturbative phase
is pushed out of the  system with increasing particle density. Even
for large particle densities compared to nuclear matter constituent
quark density, the systems show a heterogeneous structure rather than
a homogeneous transition. This situation resembles a typical percolation
transition from a hadron to a quark gas. At low particle densities the
energy scaling with the 
particle density is  characteristic for a system of strings whose
energy scales with the size of the strings. By increasing the particle
density, the flux tubes or quark bags start to overlap and to melt
into each other. The transition region between isolated hadronic
objects and a single large super-cluster, where the quarks behave as
deconfined particles, is found to be between $n = 1 \,
\text{fm}^{-3}$ and $n = 3 \, \text{fm}^{-3}$. The critical energy
density of the transition to the  deconfined phase is given between
$\varepsilon = 1.0\text{GeV}/\text{fm}^3$ and $\varepsilon =
1.3\text{GeV}/\text{fm}^3$. This transition shall be examined in the
future more carefully in a more detailed percolation analysis. There
one could measure for example the onset of the super-cluster, or the
distribution of the bag size with respect to its volume or to the
number of particles belonging to it, all as a function of the particle
density. 

\begin{acknowledgments}
  G.~M. thanks H.~St\"ocker for discussions and continuous support.
\end{acknowledgments}

\bibliographystyle{apsrev}
\bibliography{bag,cdm,dual_CSC_GL,gluon_conf,hadron_potential,numeric}

\end{document}